\documentclass[useAMS,usenatbib,usegraphicx]{mn2e}
\newcommand  \kms      {\ifmmode {\rm km\,s}^{-1} \else km\,s$^{-1}$\fi}
\newcommand  \cc       {\hbox{cm$^{-3}$}}
\newcommand  \cmii     {\hbox{cm$^{-2}$}}
\newcommand  \ergs     {\ifmmode {\rm ergs\,s}^{-1} \else ergs s$^{-1}$\fi}
\newcommand  \ergcms   {\ifmmode {\rm ergs\,cm}^{-2}\,{\rm s}^{-1}
                        \else ergs\,cm$^{-2}$\,s$^{-1}$\fi}
\newcommand  \ergcmsA {\ifmmode{\rm ergs\,cm}^{-2}\,{\rm s}^{-1}\,{\rm\AA}^{-1}
                        \else ergs\,cm$^{-2}$\,s$^{-1}$\,\AA$^{-1}$\fi}
\newcommand \ergcmsHz {\ifmmode{\rm ergs\,cm}^{-2}\,{\rm s}^{-1}\,{\rm Hz}^{-1}
                        \else ergs\,cm$^{-2}$\,s$^{-1}$\,Hz$^{-1}$\fi}
\newcommand  \phcms    {\ifmmode {\rm ph\,cm}^{-2}\,{\rm s}^{-1}
                        \else ,ph\,cm$^{-2}$\,s$^{-1}$\fi}
\newcommand  \phcmsA   {\ifmmode {\rm ph\,cm}^{-2}\,{\rm s}^{-1}\,{\rm\AA}^{-1}
                        \else ph\,cm$^{-2}$\,s$^{-1}$\,\AA$^{-1}$\fi}

\newcommand{\MBH}{$M_{\rm BH}$}

\def\micron{\ifmmode \mu{\rm m} \else $\mu$m\fi}
\def\kms{\ifmmode {\rm km\,s}^{-1} \else km\,s$^{-1}$\fi}
\def\Hubble{\ifmmode {\rm km\,s}^{-1}\,{\rm Mpc}^{-1}
        \else km\,s$^{-1}$\,Mpc$^{-1}$\fi}
\def\ergsec{\ifmmode {\rm ergs\;s}^{-1} \else ergs s$^{-1}$\fi}
\def\ergscm{\ifmmode {\rm ergs\,s}^{-1}\,{\rm cm}^{-2}
          \else ergs\,s$^{-1}$\,cm$^{-2}$\fi}
\def\ergscmA{\ifmmode {\rm ergs\,s}^{-1}\,{\rm cm}^{-2}\,{\rm \AA}^{-1}
          \else ergs\,s$^{-1}$\,cm$^{-2}$\,\AA$^{-1}$\fi}
\def\ergscmHz{\ifmmode {\rm ergs\,s}^{-1}\,{\rm cm}^{-2}\,{\rm Hz}^{-1}
          \else ergs\,s$^{-1}$\,cm$^{-2}$\,Hz$^{-1}$\fi}
\def\Msun{\ifmmode M_{\odot} \else $M_{\odot}$\fi}
\def\Lsun{\ifmmode L_{\odot} \else $L_{\odot}$\fi}
\def\qo{\ifmmode q_{0} \else $q_{0}$\fi}
\def\Ho{\ifmmode H_{0} \else $H_{0}$\fi}
\def\ho{\ifmmode h_{0} \else $h_{0}$\fi}
\def\qo{\ifmmode q_{0} \else $q_{0}$\fi}
\def\ao{\ifmmode a_{0} \else $a_{0}$\fi}
\def\to{\ifmmode t_{0} \else $t_{0}$\fi}

\def\Halpha{\ifmmode {\rm H}\alpha \else H$\alpha$\fi}
\def\Hbeta{\ifmmode {\rm H}\beta \else H$\beta$\fi}
\def\hb{\ifmmode {\rm H}\beta \else H$\beta$\fi}
\def\ha{\ifmmode {\rm H}\alpha \else H$\alpha$\fi}
\def\Hgamma{\ifmmode {\rm H}\gamma \else H$\gamma$\fi}
\def\Hdelta{\ifmmode {\rm H}\delta \else H$\delta$\fi}
\def\Lya{\ifmmode {\rm Ly}\alpha \else Ly$\alpha$\fi}
\def\Lyb{\ifmmode {\rm Ly}\beta \else Ly$\beta$\fi}
\def\hi{\ifmmode \mbox{{\rm H}\,{\sc i}} \else H\,{\sc i}\fi}

\def\ciii{\ifmmode {\rm C}\,{\sc iii} \else C\,{\sc iii}\fi}

\def\oiiihb{[O\,{\sc iii}]/H$_{\beta}$}
\def\oioiii{[O\,{\sc i}]/[O\,{\sc iii}]}
\def\oiha{[O\,{\sc i}]/H$_{\alpha}$}
\def\niiha{[N\,{\sc ii}]/H$_{\alpha}$}
\def\hahb{H$_{\alpha}$/H$_{\beta}$}
\def\lnii_ha{L([N\,{\sc ii}])/L(H$_{\alpha}$)}
\def\lniii_ha{L([N\,{\sc iii}])/L(H$_{\alpha}$)}

\def\loiii_bol{L([O\,{\sc iii}])/L$_{\rm bol}$}
\def\loiiim{L([O\,{\sc iii}])/M$_{\rm BH}$}
\def\loi_bol{L([O\,{\sc i}])/L$_{\rm bol}$}
\def\lhb_bol{L(H$_{\beta}$)/L$_{\rm bol}$}

\def\oiii{[O\,{\sc iii}]\,$\lambda5007$}
\def\oi{[O\,{\sc i}]\,$\lambda6300$}
\def\nii{[N\,{\sc ii}]\,$\lambda6584$}

\def\o5007{[O\,{\sc iii}]\,$\lambda5007$}

\def\ne212m {[Ne\,{\sc ii}]\,$12.8 \mu m$}

\def \Lop{$L_{5100}$}
\def \L5100{$L_{5100}$}
\def \Lbol{${L_{bol}}$}

\def \Lsf{${L_{SF}}$}
\def \LSF{${L_{SF}}$}
\def \Ledd{$L/L_{Edd}$}
\def \D4000{${\rm D_n4000}$}
\def \d4000{${\rm D_n4000}$}

\def  \kms         {\hbox{km s$^{-1}$}}          % kilometers per sec
\def  \ergs        {\hbox{ergs s$^{-1}$}}              % erg/sec
\def  \cc          {\hbox{cm$^{-3}$}}

\def  \cmii        {\hbox{cm$^{-2}$}}
\def  \La          {\ifmmode {\rm Ly}\alpha \else Ly$\alpha$\fi}
\def  \Ka          {\ifmmode {\rm K}\alpha \else K$\alpha$\fi}
\def  \Lb          {\ifmmode {\rm L}\beta \else L$\beta$\fi}
\def  \Ha          {\ifmmode {\rm H}\alpha \else H$\alpha$\fi}
\def  \Hb          {\ifmmode {\rm H}\beta \else H$\beta$\fi}
\def  \Pa          {\ifmmode {\rm P}\alpha \else P$\alpha$\fi}
\def  \CIIIb       {\ifmmode {\rm C}\,{\sc iii]}\,\lambda1909
                     \else C\,{\sc iii]}\,$\lambda1909$\fi}
\def  \CIV         {\ifmmode {\rm C}\,{\sc iv}\,\lambda1549
                     \else C\,{\sc iv}\,$\lambda1549$\fi}
\def  \MgII         {\ifmmode {\rm Mg}\,{\sc ii}\,\lambda2798
                     \else Mg\,{\sc ii}\,$\lambda2798$\fi}
\def  \OVI         {\ifmmode {\rm O}\,{\sc vi}\,\lambda1035
x
                     \else O\,{\sc vi}\,$\lambda1035$\fi}
\def \apj {ApJ}
\def \aj {AJ}
\def \apjl {ApJL}
\def \apjs {ApJS}
\def \pasp {PASP}
\def \mnras{MNRAS}
\def \aap   {A\&A}
\def \araa {ARA\&A}
\title[Accretion and star formation rates in type-II AGN]
{Accretion and star formation rates in low redshift type-II active galactic nuclei}
\author[Hagai Netzer]{Hagai Netzer$^{1}$\thanks{E-mail:
netzer@wise.tau.ac.il}\\
$^{1}$School of Physics and Astronomy, Tel Aviv University, Tel Aviv 69978, Israel\\
}

\pagerange{\pageref{firstpage}--\pageref{lastpage}} \pubyear{2002}

\def\LaTeX{L\kern-.36em\raise.3ex\hbox{a}\kern-.15em
    T\kern-.1667em\lower.7ex\hbox{E}\kern-.125emX}

\begin{document}
%%%%%%%%%%%%%%%%%%%%%%%%%%

\maketitle
\label{firstpage}

%\maketitle

%\title{Accretion and star formation rates in low 
%redshift type-II active galactic nuclei}
%\author{
%Hagai Netzer,\altaffilmark{1}
%}
%\altaffiltext{1} {School of Physics and Astronomy and the Wise
%  Observatory, The Raymond and Beverly Sackler Faculty of Exact
%  Sciences, Tel-Aviv University, Tel-Aviv 69978, Israel}
%

\begin{abstract}
Accretion and star formation (SF) rates in low redshift SDSS type-II active galactic nuclei (AGN) 
are critically evaluated. Comparison with photoionization models indicates that bolometric luminosity
(\Lbol) estimates  based on L(\oiii) severely underestimate
\Lbol\ in low ionization sources such as LINERs. An alternative method based on L(\hb) is less sensitive to
ionization level and a novel method, based on a combination of L(\oiii) and L(\oi), is perhaps the best. Using this method 
I show that low ionization AGN are accreting faster than assumed until now. 
Significant related other findings are:
1. Any type-II AGN property related to the black hole (BH) mass is more reliably obtained by removing blue galaxies
from the sample. 
 2. Seyfert 2s and LINER 2s form a continuous sequence of \Ledd\ with no indication for a change in
 accretion mechanism, or mode of mass supply. 
 There are  very few, if any, LINERs in all type-I samples which results in a much narrower \Ledd\ distribution
compared with type-II samples.
 3. There is a strong correlation between SF luminosity, \Lsf, and \Lbol\ over more than five orders of magnitude in luminosity.
This leads to a simple relationship between bulge and BH growth rates, $g(bulge)/g(BH)\propto L_{bol}^{-0.2}$, where
 $g(bulge)/g(BH) \simeq 115$ for  \Lbol=10$^{42}$ \ergs.
 Seyfert 2s and LINER 2s follow the same \Lsf-\Lbol\ correlation for all sources with a stellar age indicator, \D4000, smaller
than 1.8. This suggests that a similar fraction of SF gas  finds its way to the center in all AGN.
 4. \Lbol, \Lsf, \Ledd\ and the specific SF rate follow \D4000\ in  a similar way.
\end{abstract}


\section{Introduction}
Black hole (BH) masses in thousands of type-I active galactic nuclei (AGN) can now be
obtained from optical-UV spectroscopy. 
The method is based on the known relationship between the luminosity of the
non-stellar continuum at some wavelength (e.g. 5100\AA) and the mean broad line region (BLR) size derived from reverberation mapping   
(RM; e.g. Kaspi et al. 2000; Kaspi et al. 2005; Vestergaard and Peterson 2006; Bentz et al. 2009). This size
is combined with a mean gas velocity estimated from  
broad emission line profiles (e.g. \Hb), and the virial assumption about the gas motion, to obtain \MBH.
The accuracy of the method has been discussed in various papers (e.g. Bentz and Peterson 2006) and is estimated 
to be a factor of about 2. 
Complications due to the effect of radiation pressure force on the motion of the BLR gas (Marconi et al. 2008)
are probably not very important, at least for low redshift low luminosity AGN
(Netzer 2009, hereafter N09; see also Marconi et al. 2009). Estimates of the normalized accretion rate, \Ledd, 
are obtained by combining the derived \MBH\ with estimates of the bolometric luminosity, \Lbol.

Different methods are required to obtain mass and accretion rates in type-II AGN, where the 
non-stellar continuum is not directly observed. In low redshift type-II
sources situated in bulge dominated hosts, the stellar velocity dispersion in the bulge of the galaxy, $\sigma_*$,
can be transformed to BH mass (e.g. Tremaine et al. 2002). The bolometric luminosity
is usually estimated from known relationships between \Lbol\
 and the luminosity of certain narrow emission lines.
The assumption is that the line luminosity represents the same fraction of \Lbol\ 
 in all AGN. A line that is commonly
used is \oiii\ and the scaling of its luminosity to \Lbol\ is obtained from observations of type-I
sources, where both the line and the non-stellar continuum are directly observed. 
Such scaling is discussed and explained in several papers, e.g. Heckman et al. (2004), Netzer et al. (2006), 
Kewley et al. (2006; hereafter K06), Kauffmann and Heckman (2009; hereafter KH09) and N09.
A related method which is less sensitive to reddening correction is based on the luminosity of the
mid-IR line $[OIV]\,25.9 \mu$m (see discussion and suggested conversion factors in Dasyra et al. 2008).
Unfortunately, the number of sources with such mid-IR measurements is very small. Yet another
 method is based on the luminosity of the hard X-ray continuum which is directly observed in
most type-I and type-II AGN. While potentially insensitive to reddening and other complications, the exact bolometric
correction factor required to convert the 2--10 keV luminosity to \Lbol\ depends on the global SED and 
is still debatable (see e.g. the differences between Marconi et al. 2004 and Vasudevan and Fabian 2007).

The emission line scaling  method was used by Heckman et al. (2004)  to investigate mass and accretion rates in 
a large sample of data release one (DR1) Sloan digital sky survey (SDSS; York et al. 2000) type-II AGN. L(\oiii) used in this case 
included no correction for reddening and the  conversion assumed  \Lbol$\simeq 3500$L(\oiii).
The more recent KH09 analysis is based on reddening corrected values of L(\oiii).
This method reduces the scatter due to reddening and accounts, more reliably, for known differences in reddening between
galaxies of different stellar populations (K03, K06). 
The reddening-corrected scaling adopted by KH09 for all type-II sources 
is \Lbol$= 600$L(\oiii).

This paper addresses BH accretion rates, star formation (SF) and SF rates (SFR) in low redshift type-II AGN.
The aims are  to test various claims about the accretion mechanism in LINERs, the dependence of 
BH accretion on host galaxy properties, and the correlation between \Lbol\ and
SF luminosity (\Lsf). This requires  re-evaluation of the methods used to estimate \Lbol\ in type-II AGN which is 
presented in \S2. A new, improved \Lbol\ indicator is used, in \S3, to re-evaluate \MBH\ and \Ledd\ in
LINERs and in Seyfert 2s. I also use the same
indicator to test various AGN and SF correlations. \S4 summarizes the main results of the paper.
Throughout the paper I assume standard cosmology with 
H$_0$=70 km/sec/Mpc, $\Omega_m=0.3$ and $\Omega_{\Lambda}=0.7$.

\section{L(\oiii) L(\oi) and L(\hb) as bolometric luminosity indicators in type-II AGN}

\subsection{Theoretical line and continuum luminosity ratios}

The current analysis applies to narrow emission lines in galaxies that contain both an
active AGN with a narrow emission line region (NLR), and starburst (SB) ionized
gas. The term SF is perhaps more appropriate to discuss the various processes addressed below but SB has been used in the past to describe such galaxies and I keep to this terminology when discussing those issues where the name SB is commonly used.
While the NLR properties have been studied, extensively, observationally
and theoretically, the combination of AGN and SB excited gas in the same
host presents a real challenge, especially in sources observed with a large entrance
aperture. A main tool for distinguishing the two types of excitation is based on
BPT (Baldwin, Phillip and Terlevich 1981) line ratio diagrams where AGN and SB regions are well 
separated.
Detailed explanation, and critical discussion of such methods can be found in numerous papers including
Veilleux and Osterbrock and (1987), Kauffmann et al. (2003a, hereafter K03), K06, 
Brinchmann et al. (2004, hereafter B04),
and Groves et al. (2006a; 2006b).

Photoionization by a non-stellar radiation field can explain the large observed range of ionization
and excitation conditions in AGN. In particular, it can explain the L(\oiii)/L(\hb)
(hereafter \oiiihb) line ratio in the highest ionization
type-II Seyfert galaxies (\oiiihb$\sim 10$) as well as in type-II LINERs (\oiiihb$\sim 1$).
The differences are attributed to changes in the ionization parameter of the line emitting gas 
(e.g. Ferland and Netzer 1983; Netzer 1990; Groves et al 2004).
Models invoking shock excitation have also been considered (see reviews in  
Maoz 2007; Ho 2008).
Here I consider photoionization as the only ionization and excitation mechanism for the 
AGN gas. I use S2 to refer to high ionization Seyfert 2s, L2 to describe type-II LINERs and AGN
to describe the two classes together. 

Assuming ionization by the AGN continuum, it is straightforward to calculate the narrow emission line spectrum
using standard photoionization models. Realistic models must take into account the expected range of
density, column density, composition, dust content, the spectral energy distribution (SED) of the central source
 and the spatial distribution of the gas. 
 Examples are the locally optimally emitting clouds (LOC) model of Ferguson et al. (1997), the multi-cloud model of Kaspi and Netzer (1999)
 and the constant pressure model of Groves et al. (2004).
 However, the calculations of L(\oiii),
 L(\hb) and L(\oi) are relatively simple
 and single cloud models represent well the mean conditions in most NLR models. 
This has been illustrated in several recent papers such as 
Baskin and Laor (2005) and Mel{\'e}ndez et al. (2008).
  
New single cloud models have been calculated to investigate the dependence 
of the narrow line luminosities on various parameters, in particular the continuum  
 SED, the ionization parameter, the gas composition and
the dust content. The calculations were  performed with the code ION which is a standard, well 
tested photoionization code suitable for calculating NLR and BLR AGN models (Kaspi and Netzer 1999; Netzer 2006
and references therein).
Regarding SEDs, two different possibilities ought to be considered. The first is a ``standard'' SED that 
represents high ionization AGN and assumed to
retain its shape in phases of low and high accretion rates. Given the unobserved
far-UV continuum, the range of possible shapes is large and various components must be
considered. These can be described by: 
1. The shape of the 0.1--1$\mu$m continuum. 
2. The bolometric correction factor, BC, which specifies the ratio of
the total luminosity to the optical continuum luminosity. The BC is not directly observed and its value is  estimated from analysis
of various emission lines. 
3. The normalization of the optical and X-ray fluxes, e.g.
the value of $\alpha_{ox}$. 
4. The shape of the X-ray continuum. 
 
\begin{figure}
\includegraphics[width=84mm]{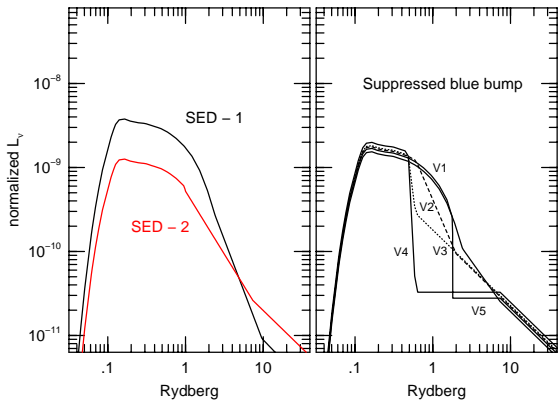}
%\includegraphics[width=80mm]{mongo_sed_pl15_model27.ps}
%\plotone{mongo_sed_pl15_model27.ps}
\caption{Left panel: Two assumed SEDs representing the range of continuum shapes
in AGN. The top curve (SED-1) is made of a disk-like optical -UV
 continuum and a powerlaw $F_{\nu} \propto \nu^{-0.9}$ X-ray (E$>$ 0.1 keV)
continuum.
For this SED, BC=7 and $\alpha_{ox}=1.35$.
SED-2 has the same optical and X-ray shapes and the two are
combined with a $F_{\nu} \propto \nu^{-1.5}$ powerlaw.
Here BC=14 and $\alpha_{ox}=1.05$.
Right panel: High accretion rate SED (V1) and four assumed low accretion rate SEDs obtained by
truncating the ``blue bump'' in different ways keeping $\alpha_{ox}$ unchanged (1.05).
}
\label{fig:SED}
\end{figure}

Experimenting with a range of possible continuum shapes shows that the two that are shown in the left 
panel of Fig.~\ref{fig:SED}  bracket most realistic SEDs.
The first of these (SED-1) is a combination of a disc-like optical-UV continuum, combined with a 
$L_{\nu} \propto \nu^{-0.9}$
X-ray ($0.1<E<50$ keV) powerlaw. In this case, BC, which is defined here as \Lbol/\Lop, where
\Lop\ is $\lambda L_{\lambda}$ at 5100\AA, is 7 and $\alpha_{ox}=1.35$. Such SEDs are typical  of high
luminosity AGN. The second case (SED-2) combines the above X-ray powerlaw and the optical disc-like continuum
with a Lyman continuum
powerlaw given by $L_{\nu} \propto \nu^{-1.5}$. In this case BC=14 and $\alpha_{ox}=1.05$. 
Such values of BC and $\alpha_{ox}$ are more typical of 
low luminosity AGN.  The two
continua shown in Fig.~\ref{fig:SED} are normalized to have the same ionizing luminosity 
to emphasize the fact that SED-1 produces relatively more photons close to the Lyman edge.

The second possibility is that different phases of activity are associated with different SEDs. For example,
a luminous, fast accreting phase may be associated with an SED similar to one of the continua shown in the left panel of Fig.~\ref{fig:SED} while a less luminous, slower accreting rate phase,
 might show a different SED. This may be the result of a change in the spectral properties of the central accretion 
 disc, in particular a significant reduction in the ``blue bump'' flux.
Such a possibility is illustrated by the various curves shown in the right panel of Fig.~\ref{fig:SED}.
The top SED, marked V1, represents the high accretion
rate phase. It is similar to SED-1 except that a relatively stronger X-ray continuum, 
typical of low luminosity AGN, has been used. In this case
BC=12.4 and $\alpha_{ox}=1.1$.
The additional curves, marked V2--V5, represent several possibilities 
where the accretion disc continuum 
is significantly reduced. 
Such spectra have been speculated for LINERs and the ones shown here were
motivated by the recent work of Maoz (2007). According to Maoz (2007),
there is a strong UV (2500\AA) point-source
continuum in several nearby LINERs. In those
cases, $\alpha_{ox} \sim 1$, very similar to the values
observed in Seyfert 1 galaxies. It thus seems that either
there are no significant changes in the near UV-SED between  low and high
accretion rate phases, or else such changes 
only affect the part of the continuum between
$\sim 2500$\AA\ and $\sim 1$ keV). The various curves in the diagram illustrate several such possibilities.
The SEDs are truncated at energies from 0.5 Rydberg,
just beyond the 2500\AA\ measured point, to 1.81 Rydberg, the HeI ionization edge.
The transitions from SED-V1 to any of the other SEDs mimic several possible changes that are consistent
with both the Maoz (2007) observations and the assumption
of a reduced disc emission.

The constant density models considered here assume column density of 10$^{21.5}$ \cmii, 
enough to make the cloud optically thick to the ionizing continuum
radiation, and hydrogen number density of $n_H=10^3$ \cc, low enough to avoid strong collisional de-excitation 
of the \oiii\ line. These are quite standard in NLR models and are in agreement with
many observations (e.g. Baskin and Laor 2005; Melendez et al. 2008; Ho et al. 1997). 
I calculated various line ratios, for all SEDs, over a large range of the hydrogen ionization parameter,
$U=Q(H)/4 \pi r^2 n_H c$, where $r$ is the
cloud central source distance and $Q(H)$ the rate of emission of Lyman continuum photons.
This was done for two generic types of
models, one with ISM-type depletion and one for dust-free gas.
The gas composition was changed between one and four times solar and the depletion fraction
 is assumed to be independent of abundance.
The models shown here are all for dusty clouds which are thought to
be more appropriate for NLR conditions.
I have included several models similar to the cases discussed in Groves et al. (2004; 2006b)
where radiation pressure force operating mostly on grains, determines the internal pressure in the cloud.
Experimenting with the various parameters confirms
that $U$, the dust content and the SED shape are the most important parameters that determine
the interesting line ratios.

\begin{figure}
%\plotone{mongo_two_models.ps}
\includegraphics[width=84mm]{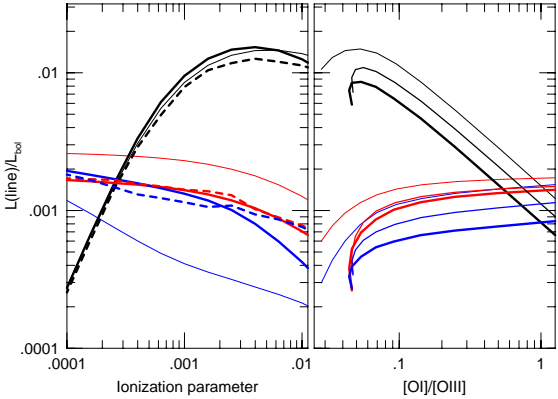}
\caption{Left panel: Calculated \loiii_bol\ (black) \loi_bol\ (blue) and \lhb_bol\ (red) ratios for various
photoionized, dusty solar composition clouds with
 full covering of the radiation source.
 Thin lines: SED-1. Thick lines: SED-2. Thick dashed line: SED-2 and radiation pressure
 dominated clouds.
Right panel: Similar calculations for the V1 continuum.
Same colour coding, same ratios as left panel but vs. \oioiii.
 Thin lines, solar abundances. Intermediate
width lines, $3 \times$solar abundances. Thick lines, $3 \times$solar, constant pressure model.
}
%Fig 2
\label{fig:calculated_oiii_hb}
\end{figure}

The results of some of the calculations are given in Fig.~\ref{fig:calculated_oiii_hb}. The figure shows
the fraction of \Lbol\ emitted by the three lines in 
question, \oiii, \oi\ and \hb. The ratios are calculated
assuming a covering factor of unity ($ \Omega = 4 \pi$).
The left panel of Fig.~\ref{fig:calculated_oiii_hb}  shows 
a noticeable difference between the behaviors of \oiii\ and \hb.
The intensity of \oiii\ depends strongly on the level of ionization and \loiii_bol\ varies by about 1.7 dex
over the relevant range of
$10^{-4} < U < 3 \times 10^{-3}$. 
 In contrast, \lhb_bol\ is basically constant (less than
 0.3 dex) over the same range which reflects the constant 
 fraction of the Lyman continuum photons absorbed by the optically thick gas.  
Such results are well known and well documented in numerous publications.
In dust-free large $U$ models (not shown here), \lhb_bol\ and \loiii_bol\ are
larger by about a factor two compared with the dusty models shown here.
Dusty gas clouds produce less \oiii\ and \hb\ photon   due to the absorption of part of 
the ionizing radiation by the
dust and the destruction of locally emitted line photons by grains. The first process is more important
in high ionization gas. 
The latter affects \hb\ more than
\oiii\ in lower ionization cases since, in this case, the \oiii\ line is formed near the illuminated face where escape out is easy.  

Comparison of the models with different SEDs in Fig.~\ref{fig:calculated_oiii_hb} shows little difference in \loiii_bol\ since the spectral shapes of the two are quite similar at high energies. This is not the case for
\lhb_bol\ because of  the larger fraction of ionizing photons just beyond the Lyman edge in SED-1 (see
Fig.\,1). All this, again, is well known from earlier calculations (e.g. Netzer and Laor 1993).

I also calculated \loi_bol\ curves and show them on the same scale. This ratio is somewhat more sensitive to $U$ than \lhb_bol\
but is much less sensitive than \loiii_bol. The differences between \oi\ and \oiii\ are
clearly seen in the right panel where I plot the same ratios as a function of \oioiii. This line
ratio varies with $U$ in a way which is different from \oiiihb. All this is the basis
for the new calibration method discussed in \S2.2.

In conclusion, the  fraction of \Lbol\ re-radiated by the narrow  \oiii\ line depends, strongly, 
on the level of ionization of the gas. The fractions of \Lbol\ re-radiated by \hb\ and \oi\ are
almost independent of $U$. 
L(\hb) seems to be the most reliable \Lbol\ indicator for pure 
AGN  excitation.

Similar line-to-continuum luminosity ratios were also calculated for the scenario illustrated in the right panel of Fig.~\ref{fig:SED}. Table~\ref{tab:line_bol} lists some of the results.
 The first row in the table lists luminosity ratios
for a high S2 accretion rate phase  
assuming the V1 SED. I fixed the value of $U$ to give \oiiihb$=10$, typical of many such sources.
The other four rows represent low accretion rate phases with the V2--V5 SEDs. Given an SED, I fixed $U$
to produce \oiiihb$\simeq 1$, typical of many L2s. 

\begin{table*}
\begin{minipage}{126mm}
\caption{\Lbol\ estimators for full covering dusty NLRs with solar composition and various SEDs 
shown in Fig.~\ref{fig:SED} }
\begin{tabular}{lcccccc}
\hline
SED &$\log U$&\oiiihb\  &\Lbol/L([O\,{\sc iii}])      & \Lbol/L(\hb) & \Lbol/L(\oi) &BC  \\
\hline
V1  & -2.58  & 10       & 74           & 776          & 1047         &12.4 \\
V2  & -3.56  & 1.06     & 851          & 891          & 933          & 11  \\
V3  & -3.57  & 1.1      & 912          & 1000         & 525          &10.1 \\
V4  & -3.49  &  1.04    & 1585         & 1660         & 501          & 9.1 \\
V5  & -3.33  & 1.1      & 617          & 692          & 708          &11.6 \\ 
\hline
\end{tabular}
\label{tab:line_bol}
\end{minipage}
\end{table*}

Examination of the table confirms that changes in SED, even as extreme as the ones considered here,
do not change much the previous conclusions. The line-to-continuum ratios change in 
a way similar to the one shown in Fig.~\ref{fig:calculated_oiii_hb} such that 
\loiii_bol\ drops much more (factors  of 10--20) compared with the drop in \loi_bol\ and \lhb_bol\
 (factors 1.5--2). The reason is that the difference in ionization potential between hydrogen and
$O^{+1}$ is small and changes in the flux of the $O^{+1}$ ionizing photons are associated with similar changes in the number of Lyman continuum photons. Most of the reduction 
in \oiiihb\ is due to changes in $U$ and cannot be explained by changes in continuum shape, at 
least not for the SEDs considered here.
 
To summarize, large changes in \oiiihb\ are associated with large changes in
\loiii_bol. In contrast, the changes in \lhb_bol\ and \loi_bol\ following a comparable
change in \oiiihb\ are much smaller even for dramatic 
changes in SED shapes like the ones considered here.

\subsection{\Lbol\ estimators}

\subsubsection{AGN samples BH mass estimates and reddening corrected line intensities}

The most reliable estimates of \loiii_bol\  and \lhb_bol\ are obtained from observations of type-I AGNs.
In such sources, \Lop, L(\oiii) and, to a lesser extent L(\hb) (due to blending with the broad \hb),
 are directly observed and luminosity dependent bolometric corrections are
known with reasonable accuracy (Marconi et al. 2004; Netzer et al. 2007; Shen et al. 2008; 
Vestergaard 2008; see summary in N09).
 This allows a direct conversion of line to continuum luminosity.

To quantify this, I used type-I data from the SDSS/DR5 sample described in Netzer and Trakhtenbrot (2007, hereafter NT07).
As explained in N09, the sample is incomplete at $z\leq 0.1$ due to the selection method of type-I AGN
in SDSS data. The completeness level, assuming a flux limited sample,  is close to 100\% at $z \geq 0.2$.
The sample contains about 9000 radio quite (RQ) type-I AGN with
  $z \leq 0.75$. The NT07 work resulted in line and continuum 
measurements, including L(\oiii), L(\hb) and \Lop, for  most of these objects. 
The number of sources in the redshift interval 0.1--0.2
is 1333 and the number of sources with reliable narrow \oiiihb\
measurements is 1172. The fitting procedure used to measure the lines, and the other details of 
the analysis are described in NT07. The bolometric
correction factor I used is somewhat different than the one used by NT07 and is taken
to be BC=$9-\log L_{44}$, where $L_{44}=$\Lop/$10^{44}$ erg\,s$^{-1}$ (see N09).

Several type-II samples are used in the present analysis. The first was extracted from  the SDSS/DR4  (Adelman-McCarthy et al., 2006) archive
which is available on the MPA URL site\footnote{www.mpa-garching.mpg.de/SDSS/DR4/}. 
The archive includes redshifts, stellar velocity
dispersion measured through the 3\arcsec\ SDSS fibre, line fluxes, the 
 D$_n$(4000) index (luminosity weighted mean stellar age) and various other properties. 
The stellar velocity dispersion was used to derive BH mass. This is a reliable procedure for elliptical and bulge
galaxies but gives poorer results for disk and pseudobulge systems. In this work I am studying mass and accretion
rate distributions in type-II AGN and I also make a comparison with type-I sources. For the first part I will try to avoid
galaxies with less reliable BH mass estimates. This requires the removal of blue galaxies from the sample and is
explained in \S3.1. For the comparison with type-I AGN, I retain the entire sample since no separation of blue and
red galaxies has been applied to the type-I SDSS sample used here.

The line fluxes for the type-II sources were re-extracted from the raw data in the archive and reddening corrected
 fluxes were obtained in two ways, using the observed \hahb\ line ratios.
The first assumes standard galactic reddening curve combined with the assumption that the intrinsic ratio is \hahb=2.86. This value is somewhat smaller than the one expected for S2s 
(3--3.1) but is a very good approximation for L2s. 
The second is aimed to duplicate the values used in K03, B04 and in Groves et al. (2006a, 2006b). Here the extinction
is given by  $A_{\lambda} \propto \lambda^{-0.7}$. Part of the motivation to use such extinction
is related to the issue of foreground vs. internal dust. Here it is only taken as one of
two possible reddening laws.
The $ \lambda^{-0.7}$ reddening law is
very flat compared with the galactic law and results with significantly larger $A_{V}$ and intrinsic 
line luminosities. For example, the median reddening for the high S/N (see below)
sample of L2s and S2s, assuming galactic reddening, is ${\rm A_V=1.07}$ and the equivalent
number for the $\lambda^{-0.7}$ law is ${ A_V=1.80}$.
Fortunately, a consistent use of the two reddening laws (see \S2.2.2)
results in only a small difference in the deduced \Lbol. 

Two additional complications related to reddening must be considered.
First, the
emission line reddening in type-II AGN is thought to be larger than
in type-I AGN. This has been discussed in various papers, e.g. Netzer et al. (2006), and Melendez et al. (2008). 
Any reddening correction based on average values obtained from type-I samples must include this effect.
Second, the average \ha/\hb\ in previous optically selected and X-ray selected samples (e.g. the Bassani et al 1999 sample)
is larger than the one measured here. This is possibly explained
by K03 who noted the large range in \ha/\hb\ and the tendency to detect the
highest values in galaxies containing more SF gas (e.g. smaller \D4000\ systems).  Several of the pre-SDSS selected samples
might preferentially pick galaxies of this type.
The present work uses reddening corrected values based on the mean observed EW(\oiii) in type-I SDSS sources. It is
therefore appropriate to use the mean value of \ha/\hb\ in the SDSS S2 sub-sample in order to choose the correct
scaling between type-I and type-II sources.

The total number of L2s and S2s in the sample considered here is about 85,000 and the maximum redshift is about 0.25.
The main purpose is to obtain the most reliable \Lbol\ by using various
emission lines. Therefore, most of the results pertain to sources  
with S/N$ \ge 3$ in the \Ha, \hb, \oiii, \nii\  and \oi\ lines. This reduces the sample size 
to about 42,000.  Out of these, about 10,000 are classified as blue galaxies (\S3.1) and are removed from part of the
analysis, as described below.

Three sub-samples were created. 
The first is defined by the \niiha\ vs. \oiiihb\ K03 criteria for separating AGN from SB galaxies. This
group is referred to here as the ``Ka03 sample''. The second and third groups are obtained by using the Ke06 division lines based on
the \oiha\ and the \niiha\ vs. \oiiihb\ line ratios (see K06). These criteria are aimed
to isolate, as much as possible, pure AGN from starburst and composite sources. The objects that are thought to best represent this group 
are in the region of the BPT diagram defined by the {\it combination} of the two K06 criteria.
There are 11,803 such sources and they are referred to here as the ``Ke06 sample''. The S2 sub-group
contains 6641 sources and the L2 group 5162 sources.
Obviously, the ratio of the populations depends on the assumed S/N since L2s have, on the average, weaker emission lines. For example, assuming S/N=1 in the Ka03 sample, I get N(L2)/N(S2)=2.5.
The samples were selected by their reddening corrected line ratios. This introduces only a small difference 
compared with the selection by uncorrected line ratios.  

An additional small sample is taken from Ho et al. (1997). These are high quality observations
of nearby sources observed through small apertures. Here the
contribution from SB regions is small and can be more easily removed.
Using the Ho et al. definitions, I choose from the tables only L2s and S2s and avoided transition
sources (e.g. those marked as L2/T2). 
BH mass estimates for most of these sources can be obtained from $\sigma_*$ measurements listed in the literature. The archive used for this purpose is LEDA
(Paturel et al. 2003)

\subsubsection{Bolometric luminosity indicators}

Every narrow emission line is a potential
\Lbol\ indicator. However, lines like \nii\ are very sensitive to the gas composition (e.g. Groves et al. 2006b)
and are not useful in this respect. This is not the case for the oxygen and the hydrogen Blamer
lines that can be used to define four such indicators:\\
{\bf 1. The \hb\ indicator}. In principle, this 
is the best indicator because of the very flat dependence of \lhb_bol\ on continuum
shape and ionization parameter (Fig.~\ref{fig:calculated_oiii_hb} and Table~\ref{tab:line_bol}). 
The normalization of  \lhb_bol\ can be obtained from type-I observations where the
optical AGN continuum is directly observed.
A major source of uncertainty is the difficult deblending of the  
relatively weak  narrow component 
from the strong, broad \hb\ line.
The fitting procedure takes this into account by imposing several criteria, such as assuming the same profile
for the narrow \oiii\
and \hb\ lines and forcing lower and upper limits on \oiiihb\ (see NT07).  
The typical  uncertainty in this procedure is estimated to be 0.2 dex.

\begin{figure}
%\plotone{mongo_L_Lbol_oiii_hb_type1_z1_z3.ps}
\includegraphics[width=84mm]{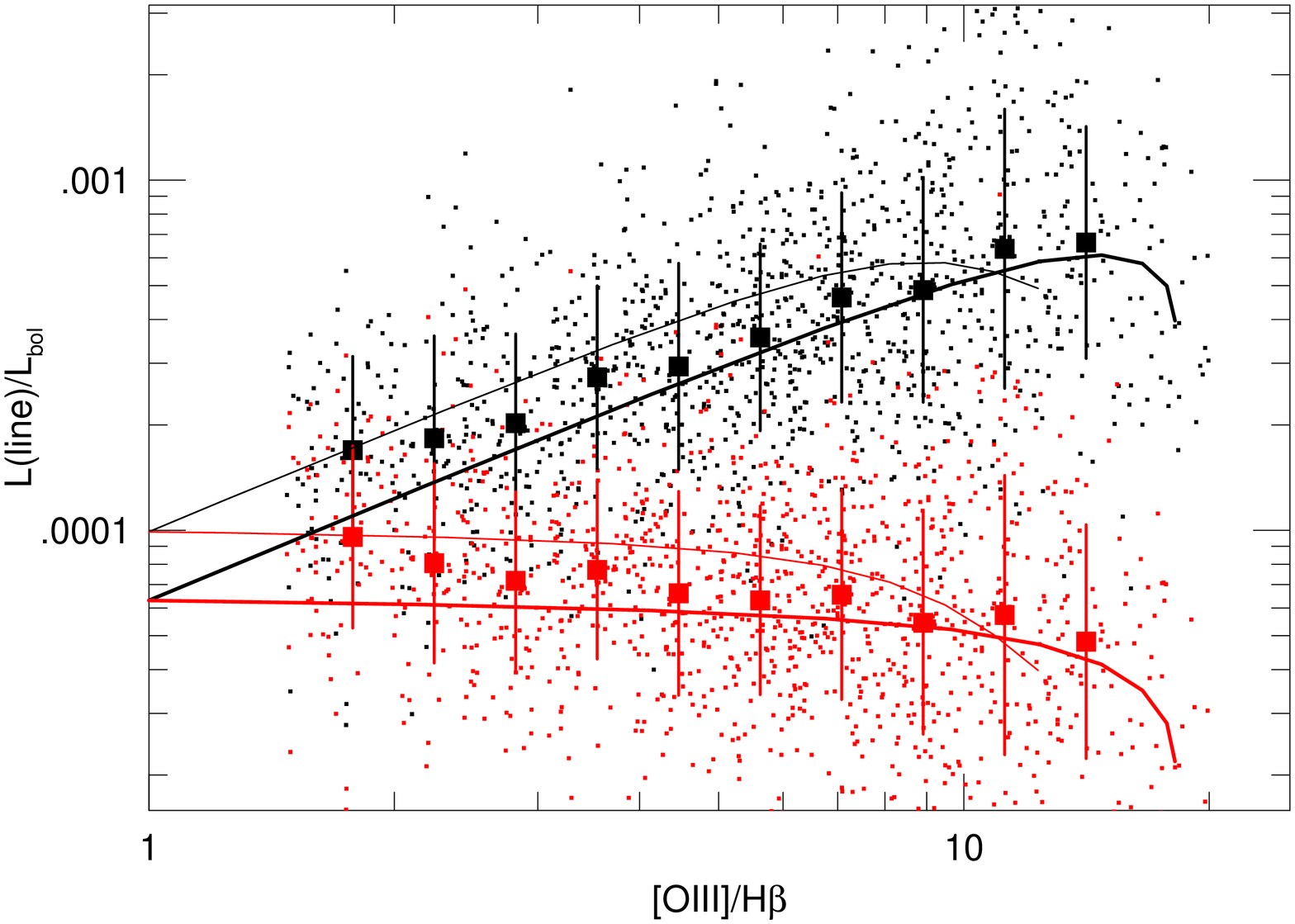}
\caption{Estimated \loiii_bol\ (black) and \lhb_bol\ (red)
in $0.1 \leq z \leq 0.2$ type-I AGN: small points represent individual objects
from the NT07 sample.
Large squares with error-bars are means and standard deviations in bins of 0.1 dex in
\oiiihb. The smooth curves are the theoretical models of
Fig.~\ref{fig:calculated_oiii_hb} (thin lines SED-1, thick lines SED-2)
where the ionization parameter is replaced by
\oiiihb. The lines are shifted down relative to Fig.~\ref{fig:calculated_oiii_hb}
by 1.4 dex. See text for more explanation.
}
%fig. 3
\label{fig:observed_type1_oiii_hb}
\end{figure}

The values of \lhb_bol\ obtained from the NT07 sample
 are shown in Fig.~\ref{fig:observed_type1_oiii_hb} as a function of
\oiiihb. Also shown are  means and standard deviations in bins of 0.1 dex in
\oiiihb. The typical standard deviation is  0.25--0.3 dex and is a combination of the  uncertainties in the
L(\hb) measurements and the scattering in covering factor (see below). 
Measurements of \loiii_bol\ in the NT07 sample are also shown. 
As expected from the photoionization calculations (Fig.~\ref{fig:calculated_oiii_hb}),
\loiii_bol\ increases almost linearly with \oiiihb\ while
\lhb_bol\ is only weakly dependent on \oiiihb.

To make a direct comparison between models and observations, I used the calculations shown in Fig.~\ref{fig:calculated_oiii_hb} to convert
ionization parameter to \oiiihb. For the case shown here I  chose the calculated  \lhb_bol\ and \loiii_bol\ in the
SED-1 and SED-2 dusty, solar composition clouds. This enables me to plot the theoretical
curves of Fig.~\ref{fig:calculated_oiii_hb} on top of the data in Fig.~\ref{fig:observed_type1_oiii_hb} 
by shifting the curves down to fit the observations. The required shift
is 1.4 dex and the agreement between model and observations is very good.
The difference between the vertical scales of Fig.~\ref{fig:calculated_oiii_hb} and Fig,~\ref{fig:observed_type1_oiii_hb}
is a manifestation of the fact that the
NLR covering fraction is much smaller than unity.
The data shown in Fig.~\ref{fig:observed_type1_oiii_hb} suggest that 
for narrow \hb\ lines in low ionization (\oiiihb$\leq 4$) type-I AGN, 
 \Lbol$\sim 14,000$L(\hb). The ratio increases smoothly to about
20,000 in high ionization parameter sources (\oiiihb$\sim 10$).

The present work uses reddening corrected line intensities yet the lines in the NT07 samples are not 
corrected for reddening. The  correction
factor can be estimated from the mean measured \hahb\ in type-II sources. As explained in K03, and
noted earlier, the amount
of reddening depends on the stellar population and is higher in smaller \D4000, large SFR galaxies. It
is therefore important to split the population into groups, e.g. S2s and L2s.
The median \hb\ reddening correction factors for S2s in the Ke06 sample are $\sim 3$ for galactic
type reddening and $\sim 5.7$ for the $\lambda^{-0.7}$ reddening law. The corresponding factors for L2s are
$\sim 1.6$ for galactic reddening and $\sim 2.2$ for the $\lambda^{-0.7}$
approximation. For comparison, the median correction factor in the Bassani et al. (1999) sample, for
galactic-type reddening, is close to 8.
The S2 factors are more appropriate since there are very few LINERs in the type-I sample 
I used for the calibration (see \S3). 
As explained, narrow emission lines in type-II sources suffer, on the average, more reddening compared with
type-I sources. The additional factor is subject to some uncertainty and is estimated
to be about 0.1--0.2 dex for \hb.

Given all these considerations, including the differences
in extinction between type-I and type-II sources, result in the following relation for
reddening corrected L(\hb) and the $\lambda^{-0.7}$ extinction  law, 
\begin{equation}
\log{L_{bol}} = \log{L(H_{\beta})} + 3.48  + max \left [ 0., 0.31 (\log \frac{[OIII]}{H_{\beta}} -0.6) \right ]  \, .
\end{equation}
For galactic reddening, the factor 3.48 should be replaced by 3.75.
The uncertainty on this ratio depends on the accuracy of the scaling shown in 
Fig.~\ref{fig:observed_type1_oiii_hb} and is estimated to be 
at least as large as the size of the error bars shown
in the diagram, i.e. about 0.3--0.4 dex.

The above expression enables me to calculate the mean NLR covering fraction. 
The estimate depends on the amount of line photon destruction by internal dust
which is already included in the photoionization calculations of Fig.~\ref{fig:calculated_oiii_hb}. 
Assuming this accounts for a factor
of 1.5--2 out of the total extinction, and a reddening correction factor of about 6 for \hb\ in S2s,
 gives a mean NLR covering fraction
of 7--15\%. For galactic type reddening, the covering fraction is smaller by a factors of 
1.5--2.  Obviously, the covering factor changes from
one object to the next which is probably the main reason for the large scatter
in \lhb_bol\ seen in Fig.~\ref{fig:observed_type1_oiii_hb}.
All these estimates are based on the assumption of little or no extinction of the optical continuum 
in type-I AGN which,
in itself, is subjected to some uncertainty (e.g. Netzer 1990).

The obvious limitation of the L(\hb) method is the difficulty in estimating the SB contribution to the observed Balmer lines, especially in large aperture observations. This has been discussed,
extensively, in K03, K06, Groves et al. (2006a),
and KH09. K06 define, for each source, a quantity which describes its distance from the SB region in the relevant BPT
diagram. Groves et al. (2006a) and KH09 suggested somewhat different procedures that serve the same
purpose. 
 The \hb\ method is more reliable in cases where the SB contributions
are small, e.g. extreme S2s with large \oiiihb\ or pure L2s. It is also the preferred method for small aperture
observations, like the Ho et al. (1997) sample.

\noindent
{\bf 2. The \oiii\ method}.
Estimates of \Lbol\ based on the observed intensity of the \oiii\ line are explained and discussed
in Heckman et al . (2004), Netzer et a. (2006), K06, N09, KH09  and several other papers. As demonstrated
in those references, for high ionization type-I AGN, the scaling factor between the observed L(\oiii)
and \Lbol\ is about 3000--3500. Netzer et al. (2006) show the clear dependence of this scaling on source
luminosity and/or redshift. This effect is not taken into account in the present work because the type-I and type-II
SDSS populations are different in terms of the fraction of low ionization (L2) sources and I
prefer to use, instead, average values for the S2s (see discussion below). Using any \oiii\ luminosity indicator includes,
therefore, another uncertainty which is related to the source luminosity.

The 
use of reddening corrected line intensities reduce the conversion factor by a factor of 3--6, depending on the extinction law used.
For the $\lambda^{-0.7}$ extinction law I find \Lbol$\simeq 600 $L(\oiii) in high ionization S2s. This is the
number used in KH09.
For galactic type extinction, \Lbol$\simeq 1100 $L(\oiii)

As explained, the major limitation of the \oiii\ method is the strong dependence 
of \loiii_bol\ on the level of ionization.
For a given \Lbol, the range in L(\oiii) is more than an an order of magnitude. In particular, a constant
 assumed \loiii_bol\ which is calibrated from S2 observations severely under-estimate
 \Lbol\ in L2s.
This issue is illustrated in the simulations described below.

\noindent
{\bf 3. The \oi\ method.}
The \oi\ line is strong enough, in many AGN and SBs, to serve as an additional \Lbol\ indicator.
The total observed range in \oiha\ is about 1.2 dex, similar to the observed range in
\oiiihb. Thus, the expected uncertainty in \loi_bol, assuming the Balmer lines are the best \Lbol\ indicators, is
large. However, the calculations (Fig.~\ref{fig:calculated_oiii_hb})
show that the \oiha\ ratio is not very sensitive
to the level of ionization of the gas, or its metallicity, and most of the range in \oiha\ 
is likely to be due to the changes in SEDs (e.g. Groves 2006a) and/or the column density of the NLR gas.
There are clearly some trends in \oiha\ that are not entirely understood.

Given the theoretical calculations, and the mean observed line ratios in the Ke06 sample,
 I estimate \Lbol$\simeq 12,000 $L(\oi) for S2s and \Lbol$\simeq 5000 $L(\oi) for L2s
 for the $\lambda^{-0.7}$ reddening law. For galactic extinction, the numbers are larger 
 by a factor of $\sim 2$.

\noindent
{\bf 4. The \oioiii\ method.}
This is a new indicator which is based on the theoretical calculations, the relative
weakness of \oi\ in SB dominated systems, and simple simulations that are discussed below.

The theoretical calculations shown in the right panel of Fig.~\ref{fig:calculated_oiii_hb}
suggest that the \oioiii\ line ratio
decreases linearly with the ionization parameter. It can  thus be used to estimate the
changes in the level of ionization 
 of the gas. This can provide the needed correction for the \oiii\ method.
The calculations also show that the above line ratio is not very sensitive to the 
gas composition, at least within the range considered here (1--4 times solar).

\begin{figure}
%\plotone{mongo_oiii_hb_oi_oiii_all_liners_nii.ps}
\includegraphics[width=84mm]{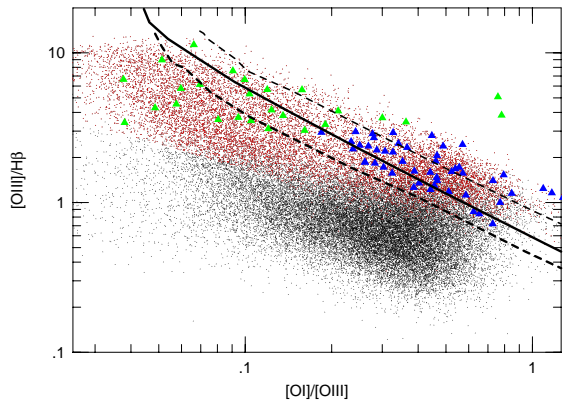}
\caption{\oiiihb\ vs. \oioiii\ for sources in the Ke06 (red points) and Ka03 (black points)
samples. The Ho et al. (1997) S2s are shown as green triangles and the L2s
as blue triangles. The black curves represent three theoretical dusty cloud models:
1. V1 SED from Fig.~\ref{fig:SED} with 3$\times$solar abundance (thick solid line).
2. V1 SED and constant pressure dominated clouds (thick dashed line).
3. SED-1 solar composition constant pressure clouds (thin dashed line).
}
%fig4
\label{fig:observed_oi_oiii_hb}
\end{figure}

To illustrate this point, I show in 
Fig.~\ref{fig:observed_oi_oiii_hb} the observed \oiiihb\ vs.  \oioiii\ 
for the  high S/N Ka03 and Ke06 
samples. The clear strong trend confirms the similar dependences of the two line ratios on the
level of ionization of the gas.
As explained, the Ke06 sample was chosen to be removed, as much as possible,
from the SB region. This sample is shown by red points and the Ka03 sample by black points.
I have also added the Ho et al. (1997)  S2s (green triangles) and L2s (blue triangles).
 As expected, these small aperture observations lie in the Ke06 zone away from
the SB region. 

To test the agreement with the theoretical calculations, I draw three lines corresponding to
three calculated models 
with various assumptions about metallicity,  SED and the role of radiation pressure (see figure caption).
All models pass through the Ke06 and the Ho et al. (1997) points verifying that
the red colored part of the diagram is indeed dominated by the AGN contribution. The curves that
represent models with different abundances also show that the
gas metallicity does not affect much the above line ratios.
The observations and the models suggest the following linear relationship
 in the AGN dominated part of the diagram,
\begin{equation}
\log \frac{[OIII]}{H_{\beta}} = -0.75 \log \frac{[OI]}{[OIII]} -0.05.
\label{eq:1}
\end{equation}
The uncertainties on the two terms, from the fit procedure only, is about 10\%.
Since \oiiihb\ is an ionization parameter
indicator, and since this parameter determines
the exact \loiii_bol\ (Fig.~\ref{fig:calculated_oiii_hb}),
the two can be combined 
to derive the following expression for the case of a  $\lambda^{-0.7}$ reddening law, 
\begin{equation}
\log L_{\rm bol} =  3.53 + 0.25 \log L([O\,{\sc III}] \lambda 5007) +0.75 \log L([OI] \lambda 6300)  \,\,  .
\end{equation}
For galactic reddening, the constant 3.53 should be replaced by 3.8.
The normalization of \Lbol\ is based on eqn.~\ref{eq:1} at the limit 
of the largest \oiiihb\ and the agreement between
theory and observations in type-I sources (Fig.~\ref{fig:calculated_oiii_hb}).
The uncertainties for individual measurements are
due mainly to the uncertainty in the slope of eqn.~\ref{eq:1}, the range in the 
NLR covering factor and the assumed bolometric correction factor,
BC. The combined  uncertainty cannot be smaller than $\sim0.4$ dex.
 
\subsubsection{Simulated BPT diagrams}

\begin{figure}
%\plotone{mongo_simulation_oiii_hb_oi_ha_agn.ps}
\includegraphics[width=84mm]{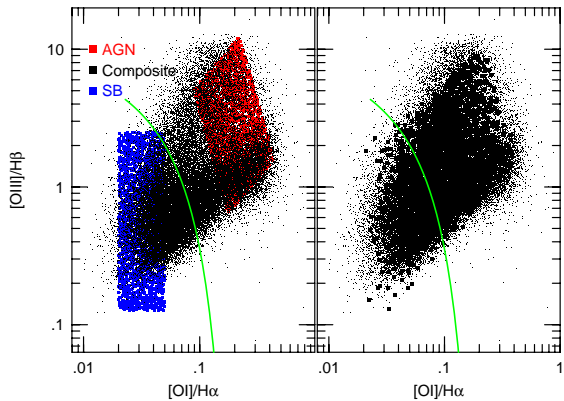}
\caption{Simulated BPT diagrams. Pure AGN are marked with red points and pure SB  with blue
points. Left panel: The assumed locations of SBs and AGN and observed sources from the Ke06 sample (small black points).
Right panel: Simulated composite sources (large black squares) and observed Ke06 sources (small black points).
The green curve marks the Ke06 AGN-SB division line.
}
%fig5
\label{fig:simulation_oiha_oiiihb}
\end{figure}

To further investigate these issues,
I ran simple simulations designed to mimic the distributions of SBs, AGN and composite sources 
in the \oiiihb\ vs. \oiha\ plane. The simulations are shown in Fig.~\ref{fig:simulation_oiha_oiiihb}.
The  starting points are  
the distributions shown in K06 Fig.\,1c and  Ho (2008) Fig.\,2. These are used to define
``pure SB'' (blue points) and ``pure AGN'' (red points) regions in the left panel of the
diagram.\footnote{Note
that B04 assigned a single location to all AGN and KH09 two
representing points, one for S2s and one for L2s. Here I use a much large
range determined by the observations} Also shown are all sources from the Ke06 sample (small black points)
and the Ke06 division line separating SB from AGN.
These simulated SBs and AGN were used to create a sample of composite sources that represent different
combinations of the two in terms of their \Lbol\ (represented by the \hb\ luminosity) and different line ratios.

I start by randomly mixing the two populations 
using their \hb\ luminosity, i.e. for a total luminosity L(\hb), a fraction $f$L(\hb) is contributed
by the AGN and a fraction $(1-f)$L(\hb) by the SB, where $f$ is distributed uniformly over the range 0--1..
I then chose, randomly, using the blue (SB) and the red (AGN) areas, values for \oiiihb\ and \oiha\ for the simulated SB and AGN.
These are combined to form a composite spectrum with composite \oiiihb\ and \oiha\ line ratios.
These simulated sources are shown as black squares in the right panel of Fig.~\ref{fig:simulation_oiha_oiiihb}.
The  composite sources are the only ones analyzed here and the 
points representing pure AGN and pure SBs in the left diagram are only shown to demonstrate their locations.
Clearly, there is a good
agreement between the areas occupied by real and simulated sources in the right panel of Fig.~\ref{fig:simulation_oiha_oiiihb}.

It is important to note that the density of the simulated sources in Fig.~\ref{fig:simulation_oiha_oiiihb} is
 {\it not} the same as the density of observed sources.  Instead, the simulations assume evenly 
spread values of $f$ and evenly  spread \oiiihb\ and \oiha\ line ratios in the SB and AGN regimes.
The diagrams emphasizes the problem of identifying the 
composite sources whose locations 
overlap with the pure SB and pure AGN regions.

\begin{figure}
%\plotone{mongo_sim_oiii_hb_dif_oiii_oi_oiii.ps}
%\plotone{mongo_sim_hb_oiiihb_dif.ps}
\includegraphics[width=84mm]{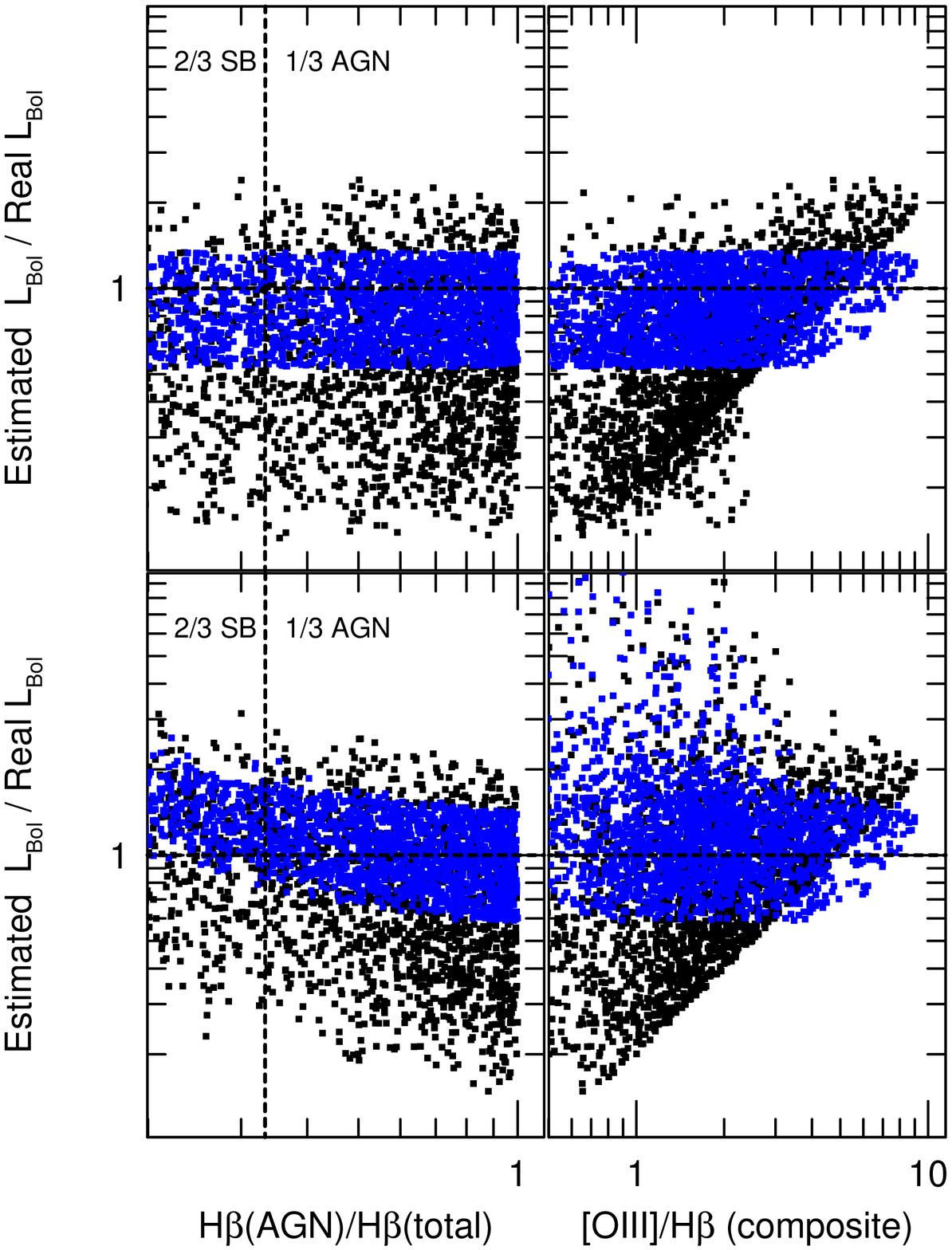}
\caption{Estimated \Lbol\ relative to the real (i.e. assumed by the simulation)
\Lbol\ using the \oiii\ (black points) and the
\oioiii\ (blue points) methods.
The upper panels assume that the SB contributions to the \oiii\ and \oi\ lines are
perfectly subtracted. The bottom panels make use of the total ``observed'' \oiii\ and \oi\
luminosities (i.e. the combined AGN-SB line luminosity).
The left panels compare the two methods for various  fraction of the total \hb\ flux
emitted by the AGN. The vertical line marks the 1/3 division line, i.e. the SB fraction in
left to the line is greater than 2/3.
}
%fig6
\label{fig:estimated_real_lbol}
\end{figure}

For each of the simulated composite spectra
I calculated the ``real'' \Lbol\ of the AGN
(i.e. the one obtained from $f$L(\hb)) 
and compared it with the ``estimated'' \Lbol, i.e. the 
one that would have been obtained by using the various \Lbol\ indicators.
Four examples are plotted in 
Fig.~\ref{fig:estimated_real_lbol}.
The top two panels compare estimated-over-real \Lbol\ using the \oiii\  (black points) and the \oioiii\ (blue points) 
indicators. The two are plotted against the 
observed \oiiihb\ and the AGN fraction in the 
composite spectrum. This simulation is done under an optimistic
assumption that the SB contribution to each of the lines can completely and accurately be removed. 
As expected, the \oiii\ indicator over-estimates the real \Lbol\ in sources with large \oiiihb\ and underestimates 
it in sources with small \oiiihb. The \oioiii\ indicator is distributed more 
uniformly, in a narrower band, with a mean 
of 1.0 and a scatter of about 0.15 dex. 
The bottom panels show similar ratios based on the
``observed'' \oiii\ and \oi\ line luminosities (i.e. the combined AGN-SB luminosity).
This results in a larger scatter but the main conclusions are unchanged. I have also tested the \oi\ method (not shown in the diagram).
The results are somewhat inferior to the \oioiii\ method but superior to the \oiii\ method. Obviously, the \oiii\ method looks somewhat better, and the \oi\ method is more biased, when the horizontal axis is replaced by \oiha.

A critical issue is the fraction of SB contribution to the \oiii\ and \oi\ lines that can be tolerated. The left panels of
 Fig.~\ref{fig:estimated_real_lbol} address this issue by showing the 1/3 division line, i.e. the location where only
a third of the \hb\ line is due to the AGN emission. At this mixing, the \oioiii\ method is still producing \Lbol\ estimates that are within
 a factor 2 of the real values. This number seems to be a practical limit when
estimating  the contamination of the \oiii\ line, and
hence the uncertainty on \Lbol, in optically selected samples.
 
\begin{figure}
%\plotone{mongo_oi_oiii_simulation.ps}
\includegraphics[width=84mm]{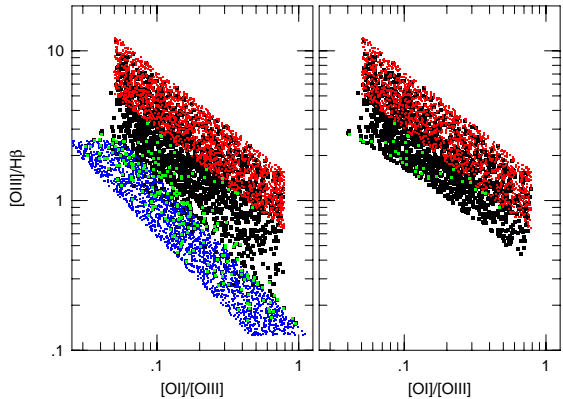}
\caption{Simulated \oiiihb\ vs. \oioiii\ diagrams obtained from the data shown
in Fig.~\ref{fig:simulation_oiha_oiiihb}.
The left panel shows the entire simulated sample and the right panel only sources
above the Ke06 division line. The green points in both diagrams are sources where
the estimated \Lbol\ obtained by the \oioiii\ method
deviates from the ``real'' (i.e. assumed in the simulation) \Lbol\
by more than a factor two.
}
\label{fig:simulation_oioiii_oiiihb}
\end{figure}

Finally I show in Fig.~\ref{fig:simulation_oioiii_oiiihb} the simulated source distribution
in the \oiiihb\ vs. \oioiii\ plane. This plot is very similar to the one obtained
from the real observations (Fig.~\ref{fig:observed_oi_oiii_hb}).
There are two parts to the diagram. The left hand side shows all composite sources (black points), pure SBs (blue)
and pure AGN (red). It also shows in green all composite sources where the \oioiii\ indicator
results in \Lbol\ which deviates by more than a factor of two from the real \Lbol\ (12\% of the sources). 
As expected, most of these sources lie close to the border line between SBs and AGN.
The right hand side is a similar diagram for sources in the Ke06 part of the diagram. The fraction of green
sources here is only 2.5\%. Thus, using the \oioiii\ estimator one expects more than 85\% of the sources in the
Ka03 AGN region to be
within a factor of two of the intrinsic \Lbol\ of the AGN. In the Ke06 region, the fraction is more than 95\%. 
Note again that the fractions quoted depend on the assumed source density across the entire plane.
Hence, they only apply to samples that are distributed with equal density over the assumed AGN and SB regions.

\subsubsection{Comparison of different \Lbol\ indicators}

\begin{figure}
%\plotone{mongo_histogram_zlt01_m1_m2_m5_m6_oiii_hb_oi.ps}
\includegraphics[width=84mm]{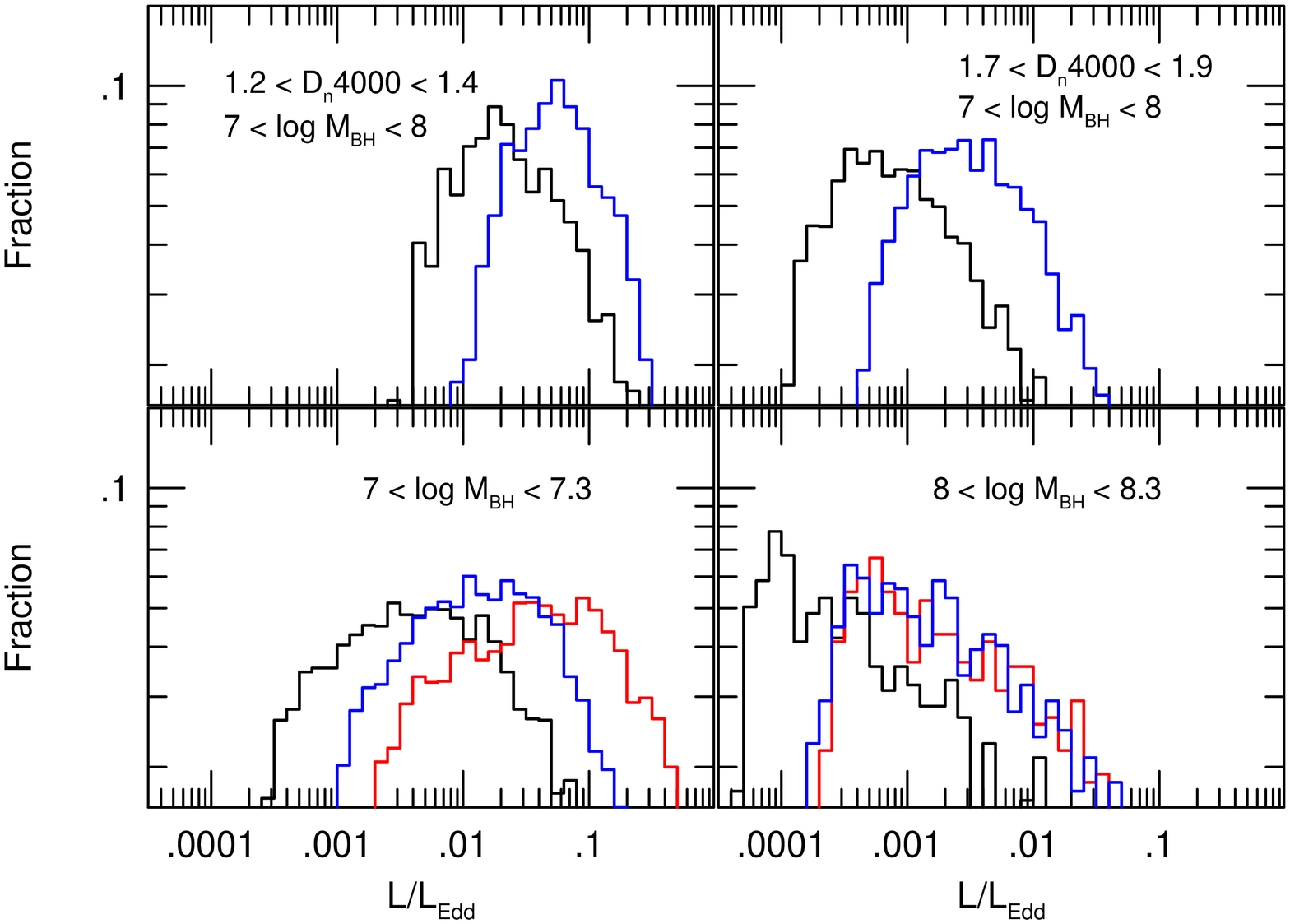}
%\plotone{mongo_histogram_zlt01_m1_m2_m5_m6_oiii_hb_oi.ps}
\caption{A comparison of the various methods for estimating \Ledd\ in several BH mass and galaxy type (\D4000)
groups for Ka03 $z < 0.1$ AGN.
The histograms show the fraction of sources in bins of 0.1 dex in \Ledd. Black curves are
values obtained with the \oiii\ method, red curves with the \hb\ method and blue curves with the\oioiii\ method.
The bottom panels show \Ledd\ distributions in two BH mass groups for all \D4000\ classes and the top panels
two \D4000\ groups with a range of \MBH, as marked.
}
\label{fig:Ledd_m1_m2_m5_m6}
\end{figure}

I compared several \Ledd\ distributions calculated with the various methods. The bottom panels of
Fig.~\ref{fig:Ledd_m1_m2_m5_m6} show the distributions in two BH mass groups, one with 
$10^{7}\leq$M(BH)$\leq 10^{7.3}$\Msun\ and one with $10^{8}\leq$M(BH)$\leq10^{8.3}$\Msun. As 
discussed in KH09, lower mass BHs are found, primarily, in hosts with younger stellar populations 
and are, on the average, faster accretors. The large BHs are, typically, in 
hosts with older stellar population and lower \Ledd. The diagram confirms this finding but shows that the
\oiii\ indicator gives consistently smaller \Ledd\
by a factor of $\sim 4$.
The bottom left panel shows also that the \hb\ indicator
over-estimates \Ledd\ in lower \MBH\ sources, because of the SB contribution to the line. The bottom right panel
shows the good agreement of the \hb\ and \oioiii\
indicators in high mass BH systems where the SB contribution is
negligible.

The upper panels of Fig.~\ref{fig:Ledd_m1_m2_m5_m6} compare  \Ledd\ 
in different groups of the stellar age indicator, \D4000\ (see e.g. K03). It suggests that 
when using the L(\oiii) indicator, \Ledd\ 
is underestimated by a factor $\sim3$ in younger systems and by a factor of $\sim 5$ in older systems
compared with the \oioiii\ method. 
The recent KH09 paper shows a powerlaw distribution of \loiiim\ in old, large \D4000\ systems. According
to KH09, this is also the distribution of \Ledd\ in those galaxies. This change in the
shape of the accretion rate distribution is explained by KH09 as due to a
different mechanism (stellar mass loss) of mass supply to the central BH.
The distribution of  \Ledd\ in the 1.7$<$\D4000$<1.9$ group shown here which is based
on the \oioiii\ method does not confirm this idea. Its shape cannot  
be fitted by a powerlaw and its mean \Ledd\ is larger by a factor
of $\sim 5$ compared with the (translated from \loiiim) \Ledd\ found by KH09. This is a manifestation of the
failure of the \oiii\ method to estimate, properly, \Lbol\ in low ionization AGN.

\begin{figure}
%\plotone{mongo_histogram_L2_S2.ps}
\includegraphics[width=84mm]{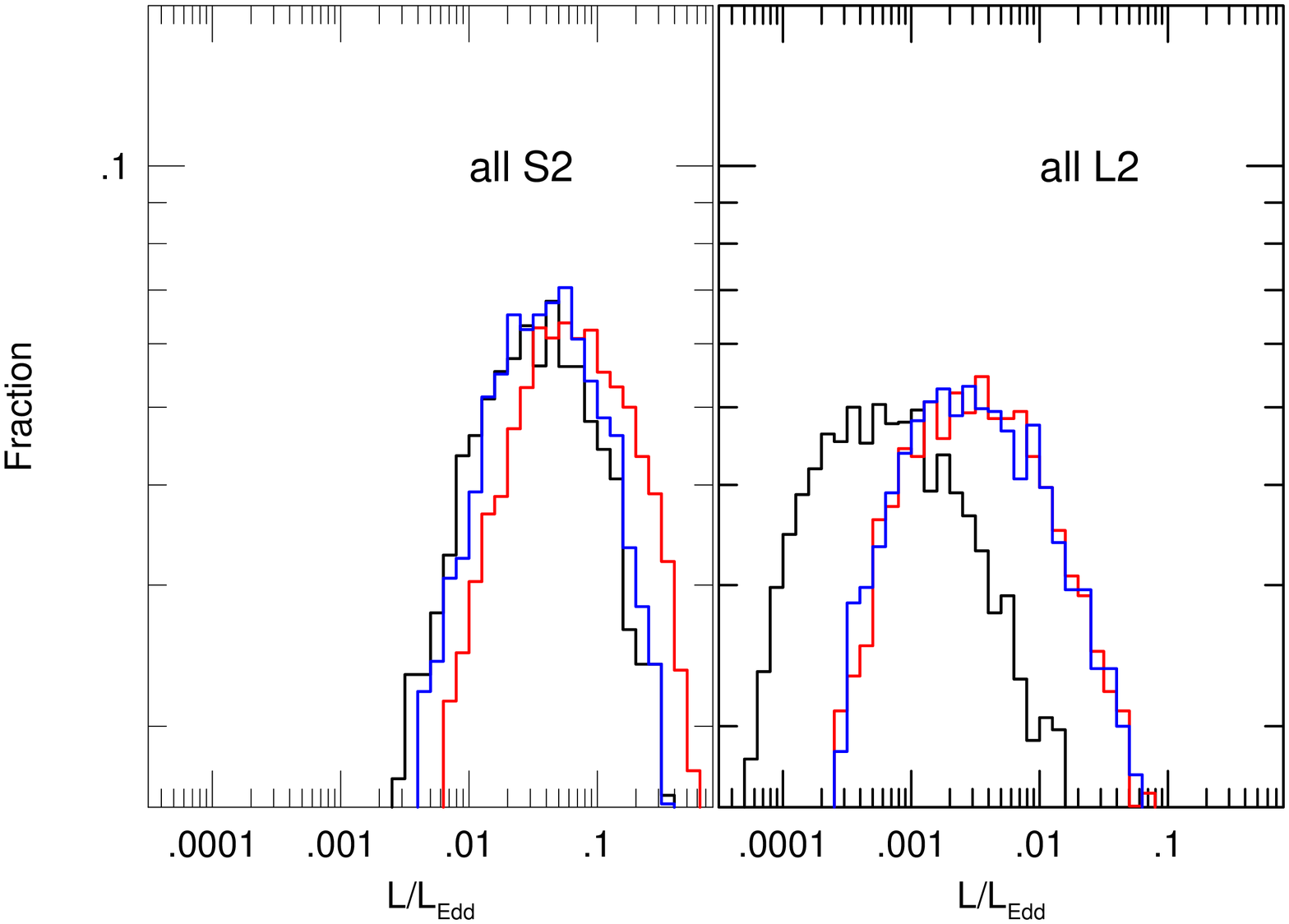}
\caption{Same as Fig.~\ref{fig:Ledd_m1_m2_m5_m6} for all Ke06 AGN at all redshifts divided into S2s and L2s.
}
\label{fig:Ledd_S2_L2}
\end{figure}

 Fig.~\ref{fig:Ledd_S2_L2} shows a comparison of \Ledd\ distributions for all S2s and L2s
 in the Ke06 sample.
 The two groups, combined, cover a large range in BH mass, $10^{6.3-9}$\Msun, and have 
 different \Ledd. 
 The diagram shows that the \oiii\ indicator is appropriate
 for the fast accreting S2s (left panel) but it  underestimates \Ledd\ in the slower accreting L2s, where the two other indicators agree very well.
 Again, the histograms are not meant to show the real distribution of \Ledd\ in all accreting
 BHs but rather to point to the most reliable \Lbol\ indicators.

 \section{Discussion}

Having established \oioiii\ as the best \Lbol\ indicator for type-II AGN, I now address two
important physical issues: the comparison of mass and accretion rates in low redshift L2s and S2s
and the correlation of \Lbol\ with SFR and the specific SFR (SSFR).
Unlike the Heckman et al. (2004) and the KH09 papers that
address the combined AGN-SB population, the present work discusses only AGN-dominated sources, those where \Lbol\ is
larger than the SF luminosity. I made no attempt to 
take into account accreting BHs in sources classified as SB galaxies. Such sources fall 
inside the SB region on the Ka03 BPT diagrams and play an important role in the KH09 analysis where, 
in several of the diagrams, they outnumber AGN.

\subsection{Black hole masses and accretion rates in low redshift S2s and L2s}

A major purpose of this work is to study mass and accretion rate distributions, and their various correlations, for L2s and S2s in the
Ka03 and Ke06 samples. Since $\sigma_*$-based estimates of BH mass are known to be reliable in bulge and elliptical galaxies, and
unreliable in disks and pseudobluges, I made an attempt to remove the less reliable estimates. The simplest approach is
to remove blue galaxies thus avoiding disks and pseudobulges (e.g. Drory \& Fisher 2007). Since $u$ and $r$ magnitudes are available for
the entire sample (SDSS ``model magnitudes''), I used the $u-r$ method described by Baldry et al.,(2004)
 to separate red and blue galaxies in the Ka03 and Ke06
samples. About 75\% of the AGN in the Ka03 sample and about 85\% in the Ke06 sample are found to be in red hosts.
 As expected, the fraction of S2s in blue galaxies is larger than their fraction in the entire population.
The method is probably too conservative since
there are  bulge galaxies that fall into the ``blue'' category, especially the larger $\sigma_*$ systems. Also, the 
border line between the groups is not as sharp as the expression suggested by Baldry et al.
 All remaining analysis that refers to BH mass and \Ledd\ 
in S2s and L2s is based on the red sub-sample only. For those cases where I compare type-I and type-II properties, I
use the entire sample since no red sub-sample is available for type-I AGN. 

\begin{figure}
%\plotone{mongo_histogram_Ledd_all_m_zlt01.ps}
\includegraphics[width=84mm]{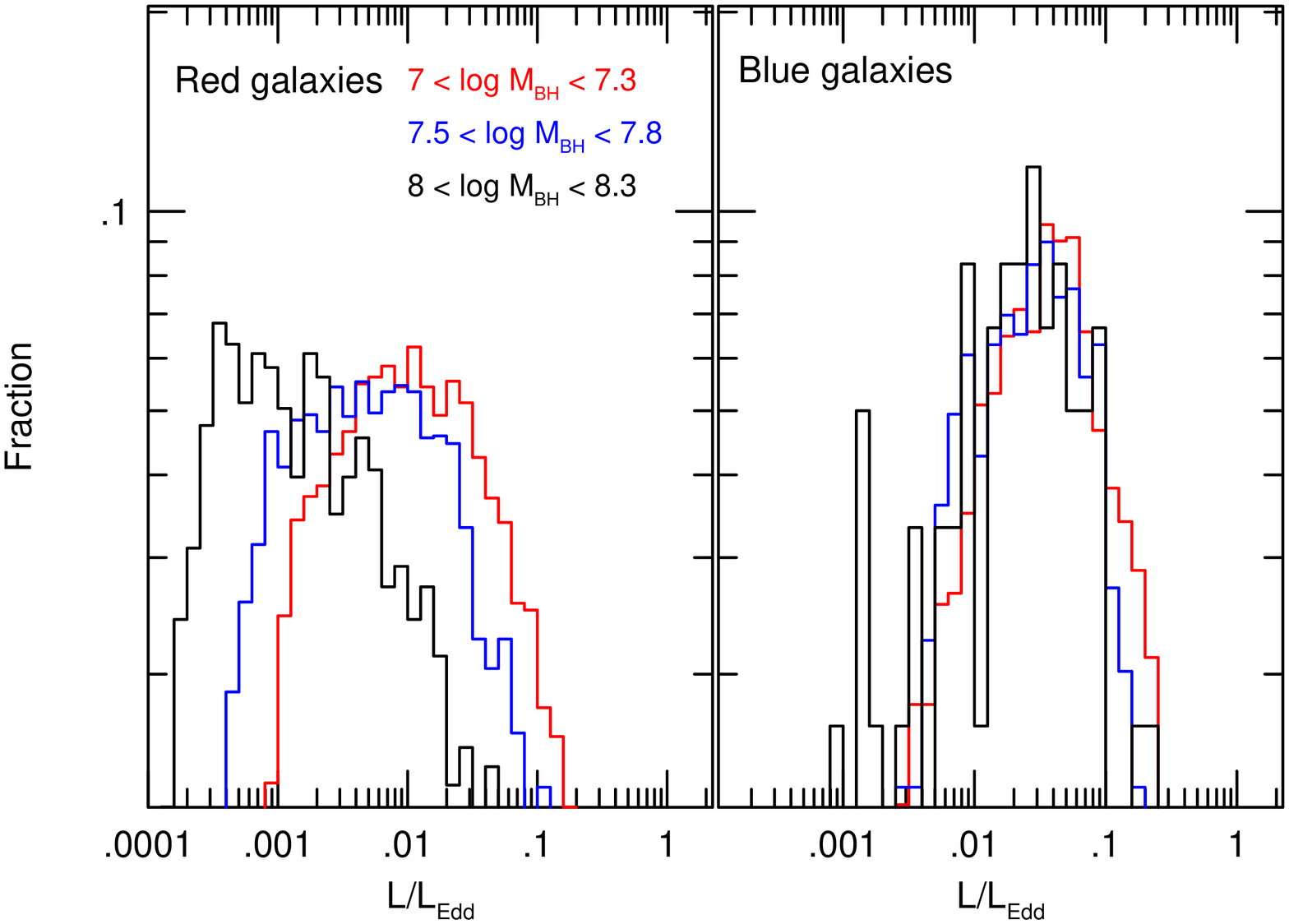}
\caption{\Ledd\ distributions of Ka03 $z \leq 0.1$ AGN with \MBH\ groups as indicated. The distributions are shown separately
for red and blue galaxies.
}
\label{fig:Ledd_all_m}
\end{figure}

Fig.~\ref{fig:Ledd_all_m} shows \Ledd\ distributions
for several \MBH\ groups in $z \leq 0.1$  blue and red hosts in the Ka03 sample.
The left panel shows the distributions in bins of 0.1 dex in \Ledd\  in the red sub-sample.
As seen, the larger BHs in the local universe are the slower accretors, similar to the behaviour in low and high redshift
type-I samples. The \Ledd\  distributions are
 broader than observed in type-I sources. This is mostly due to the lack
of L2s in type-I samples. There is no strong indication for a change in the shape of the distribution when going to larger mass BHs.
The right panel shows the same distributions for the blue sub-sample. Here there is  no difference between the mass 
groups reflecting, most probably, the unreliable \MBH\ estimates in such hosts. Mixing the two sub-samples will cause an increase in the intrinsic scatter and will tend to over-populate the larger \Ledd\ part of the diagram.

\begin{figure}
%\plotone{mongo_oioiii_Ledd_all_z_ke06.ps}
\includegraphics[width=84mm]{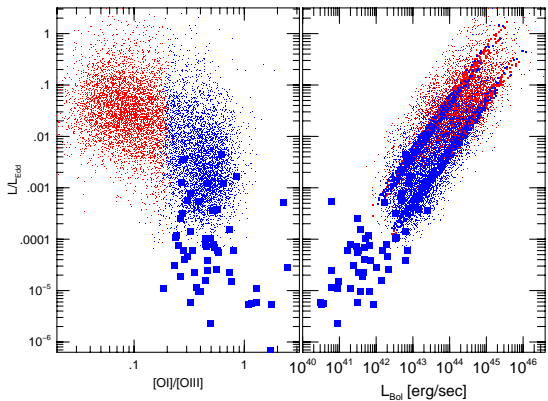}
\caption{Ke06 AGN in red galaxies divided into S2s (red points) and L2s (blue points).
The left panel shows \Ledd\ vs. \oioiii\ and the right panel \Ledd\ vs. \Lbol.
Large blue points denote L2s from the Ho et al. (1997) sample.
The two diagonal strips in the right panel are two subclasses of the S2s and L2s populations.
The top one contain AGN with $7<\log M_{BH} < 7.3$\Msun\ and the bottom one AGN
with $8<\log M_{BH} < 8.3$\Msun.
}
\label{fig:Ledd_oioiii_Lbol_S2_L2}
\end{figure}

 The dependences of \Ledd\ on  \oioiii\ and \Lbol\ for AGN in red hosts in the Ke06 sub-sample are shown in Fig.~\ref{fig:Ledd_oioiii_Lbol_S2_L2}. The
 S2s and L2s are marked with different colours to demonstrate the smooth transition between the groups.
 Plotted also are L2s from Ho et al. (1997) that probe a much lower \Lbol\ range. The left hand side
 shows that, without the Ho et al. L2s, there is a strong apparent correlation that may be interpreted as an indication that the
 \oioiii\ ratio is a good accretion rate indicator. Adding the Ho et al. L2s change this conclusion and
 illustrates the double-nature distribution of \Ledd\  with respect to \oioiii. 

 The right hand side of the diagram explores the luminosity dependence of \Ledd. As found in several earlier
 studies, \Lbol\ and \Ledd\ are strongly correlated over more than three orders of magnitudes in the SDSS sample, and over more
 than five orders of magnitude when including the Ho et al. L2s. The \Ledd\ range in the S2 class is smaller,
 $\sim 2$ orders of magnitudes. The Ho et al. sources fall on the continuation
 of the SDSS correlation. 
All \Ledd\ and \Lbol\  for L2s would drop by about a factor 5 
 if the \oiii\ would have been used instead.
\MBH\ can be considered as an additional dimension of the \Ledd\ vs. \Lbol\ correlation. This is illustrated by the two parallel
 bands where the upper stripe show L2s and S2s with
  \MBH=$10^{7-7.3}$\Msun\ and the lower one \MBH=$10^{8-8.3}$\Msun. 
 There are clear and smooth transitions between S2s and L2s
 and between all BH mass groups, over more than five orders of magnitude in \Lbol\ and no indication
  for a different sources of mass supply.

\begin{figure}
%\plotone{mongo_M_Ledd_all_z_ke06.ps}
\includegraphics[width=84mm]{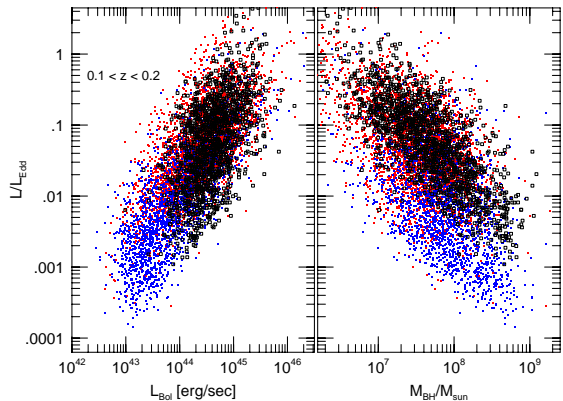}
\caption{Left: Same as right panel of Fig.~\ref{fig:Ledd_oioiii_Lbol_S2_L2} but for $0.1<z<0.2$ AGN.
Small black squares are type-I AGN from NT07. Right: the same for \Ledd\ vs. \MBH.
}
\label{fig:M_L_Ledd_typ1_type2}
\end{figure}

 Fig.~\ref{fig:M_L_Ledd_typ1_type2} shows a comparison of $0.1 < z < 0.2$ type-I and type-II SDSS AGN.
In this case I use the entire Ke06 type-II sample since the type-I group is not divided according to galaxy colour.
   The diagrams show \Ledd\ vs. \Lbol\ 
  (left panel) and \Ledd\ vs. \MBH\ (right panel).
 The division between S2s and L2s is the same as in Fig.~\ref{fig:Ledd_oioiii_Lbol_S2_L2} and the type-I sources are shown by 
 small black squares. \Lbol\ for type-Is is measured as explained in N09 (the \oiii\ method) which is
 appropriate for such objects. As seen, the type-I sources overlap, accurately, the S2s region
 in both diagrams. LINERs are completely missing from the type-I sample which results is a much
 narrower \Ledd\ range of such sources for all mass BHs. 

 The distribution of \Ledd\ in high redshift AGN samples has been studied, extensively, in several recent publications.
 Kollmeier et al. (2006), Netzer et al. (2007), Shen et al. (2007), and Gavignaud et al. (2008), used various
 selected samples, with different redshifts and depths, to address this issue. The results of the various studies 
 are quite different due to, among other things, the selection methods and the way used to estimate \MBH. 
 All those samples, as well as other low redshift type-I SDSS samples,
  clearly do not contain LINERs. The reason is the much fainter AGN continuum and broad emission lines in 
 such sources which makes their detection against the stellar
 continuum very difficult. The L2 fraction in the present low redshift sample is as large, or even larger than the S2 fraction yet all the
 above studies completely ignored this population. Obviously, there may well be fewer
 L2s at high redshifts yet it is very unlikely that such sources are completely missing. This suggests that
  the overall range of \Ledd\ in type-I AGN of all BH mass is likely to be much larger than 
 assumed so far.

\subsection{AGN and starburst correlations}

\subsubsection{Star formation in type-II AGN}
Most sources in the DR4 sample show clear indications for starburst activity
and long SF history. This has been discussed in K03, K06, Groves et al. (2006a; 2006b), 
KH09 and various other papers. Very detailed discussions, with various prescriptions of how to deduce the SFR,
are given in B04 and in Salim et al. (2007; hereafter S07).

According to B04, the SSFR in galaxies hosting type-II AGN can directly be obtained from the SDSS spectroscopy. 
This is achieved by calibrating the SB luminosity in ``pure'' SB systems against the stellar age indicator, 
\D4000, using emission line luminosities, mostly L(\ha) (see details in B04). This can then be applied to all AGN where \D4000\
has been measured to derive the SFR within the fibre. 
The above methods is subjected to large uncertainties, especially in older and redder
systems, since the calibration of the method is based on \d4000\ measurements in only SB galaxies. In older  systems,
the typical stellar age can reach several Gyrs yet most of the \ha\ emission is produced in
very young ($few \times 10^7$ yr) HII regions. This results in a big spread of SSFR for a given \D4000\ (see Fig.\,11 in B04). Estimate of the errors in using this method are given in B04 Fig.\,14. For example, at the low SF high stellar mass end, where the SSFR is below 10$^{-11}$
\Msun/yr/$M_*$, the 68 per cent confidence level approaches 1.2 dex. The error is significantly
reduced at higher SFRs. In addition, the SFRs for the entire galaxy were deduced by assuming 
that the SFR within fiber has the same dependency on colour as the SFR outside of it.
The B04 recommendation for estimating SFRs for AGN hosts is to use 
the fibre-based values for both the SSFR and $M_*$. These numbers can be found in the DR4 archive and were adopted here by using 
the median-based estimates (see general explanation in the DR4 archive).

The more recent GALEX-based work of S07 provides an independent test of the B04 method. This work uses the two
GALEX UV bands to provide UV-based SFRs for a large number of SDSS galaxies. According to S07, the agreement between the attenuation
corrected (using the Charlot and Fall 2000 two component method) 
\ha-based and UV-based SFRs is extremely good for the younger SF systems. On the other hand, large deviations can
be found in older, more massive galaxies, where the B04 method tends to over-estimate the SFR. It is however clear from their
results (e.g. Fig. 19) that some SF activity, at a level of 0.1\Msun/yr or even lower, is observed in many AGN, especially the very slow
accretors.
Inspection of the S07 results (e.g. their Figs. 3, 4 and 19) suggest that the deviations from the B04 estimates are large and the \ha-based estimates
are poor for galaxies with \d4000$>1.8$. Also, the likelihood of no SF (their ``no \Ha\'' sources) is much larger for those galaxies with total stellar mass exceeding about $10^{11.25}$\Msun. A possible reason for this over-estimation of the SFR in large mass, old stellar hosts is
the identification of a LINER excited \ha\ line as due to SF activity.
Given all these uncertainties, 
 I  chose to avoid  AGN-hosts with \d4000$>1.8$ and $M_*>10^{11.25}$ when comparing SFRs and SSFRs to \Lbol\ and \Ledd.

Having obtained \Lsf\ and SSFR for all AGN (excluding those removed by the S07-based criteria),
 I can now examine the correlations of
these properties with \Lbol\ and \Ledd. In the following I use the same \MBH\
groups and the same division into S2s and L2s as in the previous sections. The analysis
is meant to show various correlated properties 
but it is not adequate for estimating global population quantities, such as the total mass accretion onto massive BHs in the local universe.

\subsubsection{\Lsf\ and \Lbol\ correlations}

\Lbol\ in type-I AGN is readily obtained from optical and near-IR spectroscopy.
Measuring \Lsf\ in those sources is hampered by the strong non-stellar continuum and the intense
emission lines that prevent reliable measurements of the narrow Balmer lines and/or \D4000. 
The only practical way to measure \Lsf\ in luminous, high redshift sources is
based on infrared (IR) indicators, mostly PAH features in the mid-infrared (MIR) 
and cold dust emission in the far-infrared (FIR). This has been discussed in
numerous papers (see Sanders and Mirabel 1996 for the earlier work and Lutz et al. 2008 for
 references to more recent publications).

PAH-based and FIR-based estimates of \Lsf\ in low and high redshift, high luminosity type-I AGN
are described in Schweitzer et al. (2006), Netzer et al. (2007), Schweitzer et al. (2008)
 and Lutz et al. (2008). These papers describe {\it Spitzer}/IRS spectroscopy of two type-I samples. The QUEST
 sample (Schweitzer et al. 2006) contains 28 low redshift ($z \sim 0.1$) PG QSOs.
 About half of the sources show clear PAH 7.7 $\mu$m emission that is interpreted as a SF signature
 and converted to \Lsf\ using the correlation between L(PAH 7.7 $\mu$m) and L(FIR). The stacked
 spectrum of all other QUEST QSOs shows the same feature with indication of \Lsf\ which is
 only a factor of $\sim 2$ smaller for the the same AGN luminosity. Lutz et al. (2008) extended
 this study to $z=2-3$, using {\it Spitzer}/IRS observations of strong sub-mm QSOs. This provided \Lsf\
 for sources of much higher luminosity. Lutz et al. (2008) contains also estimates of \Lsf\  for additional
 high redshift AGN observed by {\it Spitzer}/IRS. The assumption in both studies is that \Lsf=$\nu L_{\nu}$ 
 at 60$\mu$m. As shown in Netzer et al. (2007) for the QUEST sample,
 and in Lutz et al. (2008) for the high redshift QSOs, there is a strong correlation between \Lsf\ obtained from
the FIR and \L5100\
 over almost four orders of magnitude in \Lbol. 
 
\begin{figure}
%\plotone{mongo_lutz_2008_fig7.ps}
\includegraphics[width=84mm]{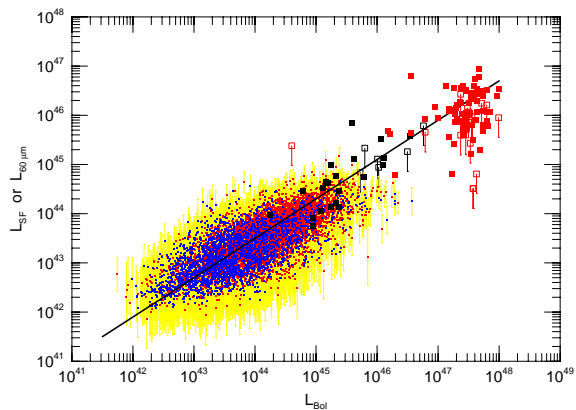}
\caption{\Lsf\ vs. \Lbol\ for AGN. The type-II Ke06 AGN are the
small points with yellow error bars. The red points are
S2s and the blue points are L2s.
QUEST QSOs from Netzer et al. (2007) are shown as large black squares 
and high redshift QSOs from Lutz et al.
(2008) as large red squares. Empty symbols with lines represent upper limits.
 The slope of the straight line is 0.8.
}
\label{fig:LSF_lutz}
\end{figure}

 I combined the present measurements of low
 redshift type-II AGN with the above L(IR)-based estimates for type-I sources to show the combined
  \Lsf-\Lbol\ diagram for  {\it AGN dominated} sources (Fig.~\ref{fig:LSF_lutz}). The error bars
on \LSF\ for individual sources  (yellow lines)
 are adopted from B04. They represent a 34 per cent confidence level.
 The low luminosity type-II AGN fall on 
 the continuation of the same correlation as the high luminosity, type-I AGN. The slope of the correlation is about 0.8 
(\Lsf$ \propto$\Lbol$^{0.8}$) and \Lsf=\Lbol\ at  about \Lbol=$10^{41}$ \ergs\ which is outside the range of the sample.

I note that the line shown here is not a fit to the data and there was no attempt to estimate the error
on its slope. Such a fit could not  be reliably obtained
not only because of the upper limits, but also because the SDSS sources, all at
the low luminosity end, outnumber the higher luminosity sources by a huge factor. It is only meant to show
that a line of this slope seems to connect all AGN-dominated sources
with real detections over a large luminosity range.
 Note also that S2s (red points) and 
 L2s (blue points) are well separated but follow the same general correlation. This suggests that \Lsf\ 
follows \Lbol\ in low as well as in high accretion rate AGN-dominated sources.

 The correlation in Fig.~\ref{fig:LSF_lutz}  may be the result of two selection effects.
 One is the omission of SF-dominated systems  above the straight line  
 and the other the absence of pure AGN below the line.
The missing sources above the line  are not due to observational 
  limitations. There are many strong SB galaxies  
   in the SDSS sample that show weak or no AGN activity and hence are
  not included in the present sample. There are also ULIRGs and SMGs with small or no AGN activity. These are also not included
in this AGN-dominated correlation.
   The bottom part of high \Lbol\ and very low \Lsf\ is more problematic and quite typical of flux limited samples. 
Given the exclusion of AGN hosts with large \d4000\ and large stellar mass, such sources are not expected to be missing at 
 low redshifts since the SDSS type-II AGN sample is thought to be complete  to about z=0.1. 
However, the high luminosity AGN samples, especially those at the highest luminosity end, are known to be 
incomplete and the fraction of missing 
 sources with very small SFR is not known. A scenario which is described below 
 gives at least one  possibility that does not require such sources.
 
\begin{figure}
% \plotone{simple_scheme.ps}
\includegraphics[width=84mm]{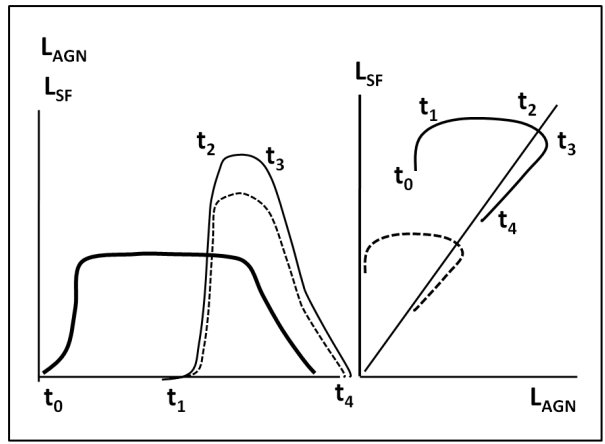}
 \caption{A schematic SF-AGN evolution sequence. The left part shows
the relevant times for the SF (thick line) and AGN (thin solid line - luminous AGN, dashed line - low luminosity AGN) evolution.
The time $t_0-t_1$ precedes the onset of the AGN accretion.
  The peak (Seyfert) AGN activity occurs between $t_2$ and $t_3$ and the
 long decay, up to $t_4$, is the LINER, low accretion rate phase.
The right part translates
 this scenario to an \Lsf\ vs. \Lbol\ relationship illustrating two tracks for law luminosity (bottom dashed line)
and high luminosity (top solid line) events (time flags are only marked for the latter). The  diagonal line on the right
represents the observed relationship from Fig.~\ref{fig:LSF_lutz}.
 }
 %Fig 14
 \label{fig:schematic_model}
\end{figure}

 The left side of Fig.~\ref{fig:schematic_model}  shows a suggested time evolution
 from pure SB (thick line) to a combined powerful SB-AGN to to a combined weak SB-AGN. This is a ``single event'' that
can occur, in principle, several times in the history of a galaxy.
The long 
 SF episode starts at time $t_0$. Some of the cold gas finds its way to the center which 
 results in the onset of BH accretion at time $t_1$. 
 The AGN rise time is short and lasts until $t_2$. This is followed by an intense AGN phase until $t_3$.
 The diminishing supply of cold gas to the SF regions, and the central BH, cause a period where the two
 fade in parallel until $t_4$. The two processes may terminate together or, alternatively, the AGN
 fading may last a little longer. In both cases, \Lsf\ and \Lbol\ are much smaller at large $t$ compared with
 their peak values. This schematic evolution resembles the S07 suggestion of a smooth sequence that begins with SF
galaxies without AGN and extends at its massive end to AGN hosts with different levels of SF. In this scenario, weak
AGN fall are associated with lower SFR relative to the peak activity of both. The strong AGN phase has  a tail that extends into the 
domain of quiescent galaxies. AGN in early type galaxies that do not show any SF activity are not shown.
The diagram shows two such scenarios with the same SFR. One for a high luminosity AGN (top solid line, large \Lbol/\Lsf) and one with 
a weaker AGN (dotted line, smaller \Lbol/Lsf).

 The above scenario is translated to an \Lsf\ vs. \Lbol\ curve in the right side of the
 diagram. Here the straight line
 represents the correlation of Fig.~\ref{fig:LSF_lutz}. Two pure SF-dominated systems with different supply of gas are shown as 
two rising and horizontal lines during  times $t_0-t_2$.
 The fading parts, $t_3-t_4$,  representing the decreasing 
 branches of both \Lsf\ and \Lbol, are shown as 
 lines going down parallel to the main correlation. The regions $t_2-t_3$ on both curves are obviously more complex. They may involve constant
 SF and BH accretion rates (i.e. one point on the curve) or periods where both \Lsf\ and \Lbol\
 increase along the main correlation,
 perhaps up to a point where \Ledd$\sim 1$. In this scenario, there are no missing objects below the correlation line
in Fig.~\ref{fig:LSF_lutz} unless the
 AGN accretion continues into the quiescent galaxy part with no SF activity. 
The left upper part of the diagram will be filled by SF galaxies that are not included in 
 Fig.~\ref{fig:LSF_lutz}. These are found in the SDSS sample and in several, high redshift LIG, ULIRG and SMG samples.

  The correlation in Fig.~\ref{fig:LSF_lutz} can be translated to a ratio between the bulge growth rate, $g(bulge)$,
  assumed to be proportional to \Lsf, and BH growth rate, $g(BH)$, assumed to be proportional to \Lbol.
  For BH radiation conversion efficiency of $\eta_{BH}$, and SF radiation conversion efficiency of
  $\eta_{SF}$, the ratio is 
  \begin{equation}
   \frac{ g(bulge)}{ g(BH)} \simeq 115 \left [ \frac{\eta_{BH}/0.1}{\eta_{SF}/7\times10^{-4} } \right ]
     \left [ \frac{ L_{Bol} }{ 10^{42}\,{\rm erg\,s^{-1}} } \right ] ^{-0.2} \,\, .
  \label{eq:g_g}
  \end{equation}
In bulge dominated systems, the time integrated ratio of these quantities must equal the local measured value
of $M_*/M_{BH}$. This number is larger by a factor of at least six
compared with the value inferred from eqn.~\ref{eq:g_g}. It suggests that even the slowest accreting BHs in AGN dominated sources, 
those at the bottom left part of Fig.~\ref{fig:LSF_lutz}, with SFR$\sim0.1$ \Msun/yr, are still growing at a relative rate 
which is about six times faster than their integrated cosmic growth rate. For the fastest accreting BHs in type-II AGN 
the ratio is larger than 20. The constant (115) in the above equation is similar to the one found by Silverman et al. (2009)
in their analysis of the BH growth rate and SFR in zCOSMOS galaxies (see their Fig. 13).
 These numbers can be translated, in a simplistic way, to the ratio of duty cycles between
BH  and SF activity. It it difficult to assign similar numbers to the high redshift population since $M_*/M_{BH}$
is not known at high redshift.  

Heckman et al. (2004) studied the $M_*/M_{BH}$ ratio in SDSS sources and found good 
agreement between the total accumulated $M_*$ and \MBH. Their population average is consistent with 
$g(bulge)/g(BH) \sim 1000$. This value should not
be compared with the one in eq.~\ref{eq:g_g}
since the present work consider only
AGN-dominated sources while Heckman et al. (2004) included SF galaxies that outnumber 
AGN in the SDSS sample by a large factor. A good idea of the ratio of the two is given in S07 where it is shown that 88\% of the local SF
occurs in SF galaxies and 11\% in galaxies that host an AGN. This number is consistent with eqn~\ref{eq:g_g}. All these factors must
be related, in a general sense, to the different duty cycles of SF galaxies and AGN.

\subsubsection{\Lsf\ \Lbol\ and stellar ages}

A possible way to test the simple theoretical scenario of Fig.~\ref{fig:schematic_model} is to use the measured \D4000\ 
as an evolution indicator. K03 and B04 discussed \D4000\ extensively showing that
 it represents the weighted age of the stellar population. An estimate of this age can be obtained from theoretical models.
 For example, the single burst model shown in Kauffmann et al. (2003b) gives stellar ages of 
 4$\times 10^8$ Mys for \D4000=1.3, $10^9$ yrs for \D4000=1.5 and 10 Gys for 
  \D4000=2. Obviously metallicity and various assumptions about the nature of the burst will change
 those numbers. 
 
\begin{figure}
%\plotone{mongo_d4000_LSF_LBOL_all_sources_zlt01_m3_m4.ps}
\includegraphics[width=84mm]{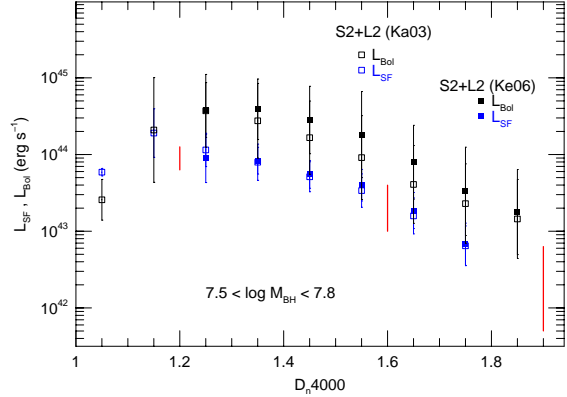}
\caption{\Lbol\ (black symbols) and \LSF\ (blue symbols) vs. \D4000\ for $7.5 < \log M_{BH} < 7.8$\Msun\
$z<0.1$ AGN in two samples: the Ka03 sample (all AGN, empty squares)
  and the Ke06 sample (full squares). 
Red vertical lines show 68 per cent confidence intervals on \LSF.
}
\label{fig:D4000_Lbol_LSF}
\end{figure}

Fig.~\ref{fig:D4000_Lbol_LSF} shows \Lbol\ and \Lsf\ vs. \D4000\ for various groups of AGN. The mean 
luminosities are calculated in intervals of 0.1 in \D4000\ and the error is represented by the standard deviation
in each bin. While the errors shown are relatively small, this is partly due to the choice of data in the
archive which picks for each \D4000\ the median of the SSFR distribution. As shown in B04 Figs.\,11 and 14,
the real distribution is much broader. To include these errors I plot in red the 68 per cent confidence
intervals on \LSF\ at various \d4000\ as given in B04. As explained, \Lsf\ in hosts with \d4000$>1.8$ are highly uncertain
and hence are not shown.
Note also that the diagram is for a given BH mass group that
contains both S2s and L2s. The S2s are the main contributes to both \Lbol\ and \Lsf\ in each \d4000\ group.

The dependence of both \Lbol\ and \Lsf\ on the age of the 
stellar population age has been known for some time (K03, S07, KH09). However, the great
similarity in the shape of the two is a new feature. It indicates, yet again, a proportionality of \Lbol\ and
\Lsf\ over a large range of stellar populations. This is another manifestation of the \Lsf-\Lbol\ correlation shown in Fig.~\ref{fig:LSF_lutz}.
The general behaviour resembles the schematic scenario of Fig.~\ref{fig:schematic_model} since \D4000\ 
is an age indicator. The drop 
at \D4000$ \leq 1.3$ is highly uncertain due to the small number of sources in this range. 

\begin{figure}
%\plotone{mongo_d4000_Ledd_SSFR_m1_m6_all_SN3.ps}
\includegraphics[width=84mm]{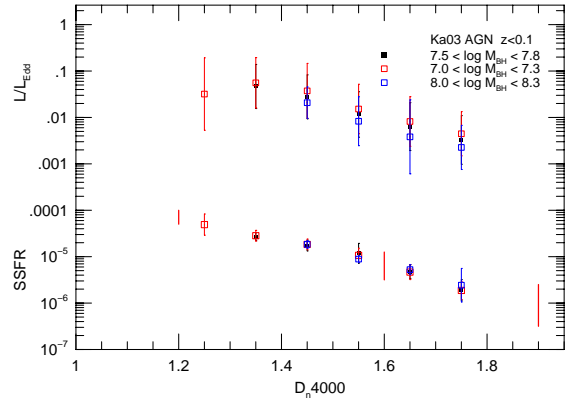}
\caption{\Ledd\ and SSFR (normalized to the same units as \Ledd) vs. \D4000\ for Ka03 $z<0.1$ AGN in red galaxies in three mass groups.
Red vertical lines show 68 per cent confidence intervals on the SSFR.
}
\label{fig:D4000_Ledd_SSFR}
\end{figure}

Division by the appropriate masses enables a 
comparison of \Ledd\ and the SSFR with \D4000. I plot these quantities in  Fig.~\ref{fig:D4000_Ledd_SSFR}
after converting the SSFRs to the same scale as \Ledd\ (i.e. multiplying
$M_*$ by the Eddington luminosity of a solar mass BH). The overall shape of the upper curve is similar to the
previous correlation. It shows a slight dependence of \Ledd\ on \MBH\ (the smaller BHs are
the fastest accretors). The lower curve is only shown for comparison since its shape simply
reflect the SSFR-\D4000\ correlation in the DR4 archive used to obtain the SFRs.
The short red lines are, again, the B04 68 per cent confidence intervals on the SSFR.
 
Returning to Fig.~\ref{fig:LSF_lutz}, and the various diagrams involving \D4000, one can propose a simple explanation 
for the observed range of \Lsf\ and \Ledd\ in a given \Lbol\ bin. The first may reflect the  
  combination of early and late type stellar populations for the same luminosity AGN and the second the combination of
 BHs with different accretion rates, all having the same \Lbol. This is a natural
 consequence of the schematic evolution between times $t_2$ and $t_4$ shown in  Fig.~\ref{fig:schematic_model}.

%In a recent paper, Chen et al. (2009) suggested that 
%The value of \Lsf\ deduced here allows a direct comparison of ...
%The slope that I obtain is about 1.4, compared with a slope of 1.8 obtained by Chen et al. 
%This would give a different history ....
 
 \section{Conclusions}
 A critical evaluation of the various ways used to estimate \Lbol\ in type-II AGN suggests that
 some of the previous methods, in particular the one based on  L(\oiii), are problematic.
 For LINERs, and for AGN in hosts with older stellar
 population and large BHs, this method under-estimates \Ledd\ by factors of $\sim 5$.
  The paper suggests several alternative
 methods for estimating \Lbol\ and use them to investigate the properties of type-I and type-II low redshift AGN from the SDSS sample. The
 main conclusions are:\\
 1. The best method for estimating \Lbol\ in type-II AGN is based on a combination of L(\oiii)
 and L(\oi). The method is appropriate for both high (S2s) and low (L2s) ionization AGN.\\
 2. BH mass and \Ledd\ distributions of S2s  resemble those of similar luminosity type-I sources.
The comparison between the two types is based on different ways of measuring \MBH\ and care must be taken in 
using the $\sigma_*$ method in blue galaxies, especially disk dominated and  pseudobulge galaxies.
Most type-II AGN are L2s with very small values of  \Ledd\ and there are 
 very few, if any, such sources in type-I samples. This results in a much narrower \Ledd\ distribution
for type-I AGN.
 \\
 3. S2s and L2s form a continuous sequence of \Ledd\ with no indication for a change in the 
  mode of mass supply. The overall range in \Ledd\ for the entire population is
 about five orders of magnitude while for S2s it is only three orders of magnitude.\\
 4. There is a clear and strong correlation between \Lsf\ and \Lbol\ over more than five orders of magnitude in luminosity
in AGN-dominated systems. The low luminosity, low redshift type-II sources fall on the same correlation as
  the more luminous, high redshift, type-I sources. There
  is no distinction between S2s and L2s all the way down to a SFR of 0.1 \Msun/yr and both follow the same \Lsf-\Lbol\ correlation.\\
 5. \Lbol, \Lsf, \Ledd\ and the SSFR all follow, in a similar way, the \D4000\ sequence up to \D4000=1.8, the limit imposed here.
 Combined with the \Lbol-\Lsf\ correlation, this may give a clue to the sequence of events that
 leads from SF to AGN activity in individual sources.

%\acknowledgments 
\section*{Acknowledgments}
I am grateful to Benny Trakhtenbrot for help in using the DR4 archive and for
useful discussions. Dan Maoz provided important insights about LINERs and their properties.
I have benefited from discussions and exchange of information with G. Kauffmann and T. Heckman.
Useful discussions with  Kristen Shapiro, Amiel Sternberg, John Kormendy, Niv Drory, Karl Gebhardt and Yan-Mei Chen are gratefully acknowledged.
I thank MPE and his director, R. Genzel, for their hospitality during several visits to their institute.
Funding for this work has been provided by the Israel Science
Foundation grant 364/07 and by the Jack Adler Chair for Extragalactic Astronomy.

\bsp
\label{lastpage}
\end{document}